\documentstyle[12pt,aps,epsf,epsfig,floats]{revtex} 
\begin{document}

\def\half {{1\over 2}}
\newcommand{\m}{\multicolumn}
\newcommand{\ba}{\begin{array}}
\newcommand{\ea}{\end{array}}
\def\be {\begin{equation}}
\def\ee {\end{equation}}
\def\3s1 {$^3S_1$}
\def\3d1 {$^3D_1$}
\def\1s0 {$^1S_0$}
\def\3p2 {$^3P_2$}
\def\5h3 {$^5H_3$}
\def\5p3 {$^5P_3$}
\def\1f3 {$^1F_3$}
\def\3p0 {$^3P_0$}
\def\3P0 {$^3P_0$}
\def\1p1 {$^1P_1$}
\def\J {\rm J}

\title{\small \rm \begin{flushright} 
\end{flushright} 
\Large \bf 
Meson-Meson Scattering 
in the Quark Model: \\
Spin Dependence and Exotic Channels \\
\vspace{0.8cm} }

\author{
T.Barnes,$^{1-4}$\thanks{email: barnes@bethe.phy.ornl.gov}
N.Black$^2$\thanks{email: nblack@nomad.phys.utk.edu}
and 
E.S.Swanson$^{5,6}$\thanks{email: swansone@pitt.edu}
}

\address{
$^1$Physics Division, 
Oak Ridge National Laboratory \\
Oak Ridge, TN 37831-6373, USA  \\  
$^2$Department of Physics and Astronomy,
University of Tennessee \\
Knoxville, TN 37996-1501, USA \\
$^3$Institut f\"ur Theoretische Kernphysik
der Universit\"at Bonn \\
Nu\ss allee 14-16,
D-53115 Bonn, Germany\\ 
$^4$Forschungszentrum J\"ulich GmbH,
Institut f\"ur Kernphysik \\
D-52425 J\"ulich, Germany\\          
$^5$Department of Physics and Astronomy,
University of Pittsburgh,  \\
Pittsburgh, PA 15260, USA \\
$^6$Jefferson Laboratory, 
12000 Jefferson Ave., 
Newport News, VA 23606, USA \\
}

\maketitle

\begin{center}
{\bf Abstract}
\end{center}

\begin{abstract}
We apply a quark interchange model to
spin-dependent and exotic meson-meson scattering. The model
includes the complete set of standard quark model forces,
including OGE spin-orbit and tensor and scalar confinement spin-orbit.
Scattering amplitudes derived assuming SHO and Coulomb plus linear plus 
hyperfine meson wavefunctions are compared.
In I=2 $\pi\pi$ we find approximate agreement with the S-wave
phase shift from threshold to 1.5~GeV, where we predict an extremum
that is supported by the data. 
Near threshold we find rapid energy dependence that may reconcile 
theoretical estimates of small scattering lengths with experimental
indications of larger ones based on extrapolation of
measurements at moderate ${k_\pi}^2$.
In PsV scattering we find that the quark-quark 
L$\cdot$S and 
T
forces map into 
L$\cdot$S and 
T
meson-meson interactions, and the P-wave L$\cdot$S force
is large.
Finally we consider scattering in J$^{PC_n}$-exotic channels, and note
that some of the ``Deck effect''
mechanisms suggested as possible nonresonant origins of the 
$\pi_1(1400)$ signal 
are not viable in this model.
\end{abstract}

\section{Introduction}

The determination of scattering
amplitudes between pairs of mesons is an interesting 
problem in strong QCD. 
It is also a complicated problem, 
because
both $q\bar q$ annihilation to $s$-channel resonances and
``nonresonant'' scattering are important effects, 
and it is often difficult to 
separate
the various contributions. However by specializing to 
annihilation-free
channels such as I=2 $\pi\pi$ and $\pi\rho$, I=3/2 K$\pi$, KN and 
NN, one may study nonresonant scattering in relative isolation.
The determination of resonance parameters, reaction mechanisms, and many
other aspects of hadron physics are complicated by the presence of 
nonresonant scattering, which is treated as an (often
poorly understood) initial-state and final-state rescattering effect.
Developing an accurate description of nonresonant scattering would 
help
clarify many other aspects of hadron physics.

A further interesting possibility is that  
sufficiently attractive
nonresonant scattering 
may lead to weakly bound hadron-hadron or multihadron states,
as does happen in nuclei and hypernuclei.  
We may also find a rich spectrum of meson-meson bound states,
the study of which will
extend nuclear physics into
the largely unexplored field of ``mesonic nuclei" or ``molecules"
\cite{WI,DSB,NT}. 

An understanding of PsPs, PsV and other meson-meson 
scattering amplitudes is also important
for the interpretation of non-QCD processes such as
nonleptonic weak decays, 
since these show evidence of important hadronic 
final state interactions. The $\Delta$I=1/2 rule
is a well known example. Similarly, a
recent study of D and D$_s$
decays to K$\bar {\rm K}\pi$ \cite{charm} found that
the 
Dalitz plots are dominated by two-meson isobars,
including $\phi\pi$, K$^*\bar {\rm K} + h.c.$ 
and K$^*_0(1430)\bar {\rm K} + h.c.$, 
and complex relative amplitudes are 
required to describe the D$^+$ Dalitz plot.
Without final state interactions one would expect
relatively real 
couplings to these final states. 

One finds a surprising variety of approaches to 
strong hadron-hadron scattering
in the literature. There 
are many studies 
using 
effective hadronic lagrangians, such as 
the ``chiral perturbation theory" description of 
the PsPs sector. 
Although this method is convenient because
it uses perturbative QFT techniques, it is incomplete in that it takes 
effective lagrangian vertex strengths from the
data; 
one 
should be able to calculate these hadronic couplings 
directly from 
quark-gluon forces. 

Second, there are studies that 
model the low energy hadron-hadron scattering mechanism, which include
the apparently
dissimilar meson exchange and quark-gluon descriptions of hadronic forces.
Meson exchange models are again attractive for their simplicity, since
they use
perturbative QFT techniques to determine scattering amplitudes.
This approach has been elaborated in greatest detail 
in models of the NN force \cite{NN_Bonn}, in which
a large number of meson exchanges is assumed. With this large parameter
space a good description of this interaction is possible, although
there is a concern that one may be parametrizing other scattering mechanisms
in addition to $t$-channel meson exchange.
Alternatively, one may calculate hadron-hadron forces directly
from the fundamental quark-gluon interaction, using quark
model hadron wavefunctions. This approach has also seen its most
detailed development in studies of the NN interaction
\cite{NN}, 
and is 
most successful in describing the short-ranged repulsive core.
Maltman and Isgur \cite{NN} also found a
physically reasonable intermediate ranged
attraction from a color van der Waals effect
in the quark-gluon
approach, which is {\it not} equivalent to the usual 
$\pi\pi$ or $\sigma$ meson exchange explanation of this force.
The quark description of hadron-hadron
interactions is complicated by the combinatorics
of matrix elements between quark bound states, but has
the advantage that 
it can easily be extended to a wide range of spin
and flavor channels through a simple change of the external hadron 
wavefunctions.

A third promising approach is to infer hadron scattering amplitudes from LGT.
To date LGT has seen little application to scattering problems because of
the difficulty of treating systems that are not in their ground states.
Estimates of the I=0 and I=2 $\pi\pi$ scattering lengths have been 
obtained by exploiting 
a theoretical relation to
finite-size effects \cite{LGT_pipi}, and more recently very
interesting results for 
nuclear physics potentials in the ${\cal BB}$ system were reported
\cite{LGT_BB}. In future it may be possible to improve 
hadron scattering models
through comparisons with similar ``LGT data''.

In this paper we are concerned with the derivation of meson-meson
scattering amplitudes from quark-gluon forces. We derive 
meson-meson scattering amplitudes at lowest
order in the quark-quark interaction, which leads to a
quark interchange model described by
``quark Born diagrams'' \cite{pipi,S_annphys}.
Since the quark-quark interaction is already well established from
hadron spectroscopy, our predictions have little parameter freedom.
In previous work we and others 
(usually assuming OGE hyperfine dominance) 
have
shown that this approach
gives a reasonably accurate description of S-wave scattering in a 
wide range of 
channels without $q\bar q$ annihilation, including
I=2 $\pi\pi$
\cite{pipi}, I=3/2
K$\pi$\cite{Kpi}, 
I=0,1 KN\cite{KN}, 
I=0,1 BB\cite{BB} (compared to LGT data),
and 
the NN repulsive core \cite{NN}. 
This approach
has also been applied 
to 
$\pi J/\psi$ 
\cite{Ros,CYW}
and other reactions
relevant to heavy ion collisions, where the experimental 
low energy cross sections
are as yet unclear.

The principal new contribution of this paper is a detailed analytical 
derivation of the meson-meson scattering 
amplitudes that follow from the
complete quark-quark interaction, including color Coulomb, linear scalar
confinement, OGE spin-spin, OGE spin-orbit, OGE tensor and linear 
spin-orbit forces. 
As a future application of these results, 
one might hope to clarify the relationship between meson exchange and
quark interchange models by a detailed comparison of the spin dependence of 
hadron-hadron scattering amplitudes,
which we expect to be sensitive to the details of the 
scattering mechanism.

Here we consider both pseudoscalar-pseudoscalar (PsPs) and  
pseudoscalar-vector (PsV) scattering. The former is a 
``standard benchmark''
for meson scattering models, because I=2 $\pi\pi$ low energy scattering
has no $s$-channel resonances and has been the subject of
many
experimental phase shift analyses.
Although we find
reasonable agreement with S-wave I=2 $\pi\pi$ scattering, 
this channel has
no spin degree of freedom, and so cannot be used to
test the characteristic spin 
dependences predicted by the quark model's OGE and linear scalar confinement
forces. 

We find in contrast that 
PsV is an excellent theoretical laboratory for
the study of spin-dependent forces, as it can accommodate both meson-meson
spin-orbit and tensor interactions. The spin-dependent
forces at the meson-meson level are closely related to the 
corresponding terms in the quark-quark interaction in our approach.
Although the study of PsV scattering is essentially a theoretical exercise
at present, these phase shifts are accessible 
experimentally, for example through measurement of the relative
S and D final state phases in $b_1\to\pi\omega$. Thus it should be possible 
to measure
PsV phase shifts 
from resonance decays
to multiamplitude PsV final states.

Before we proceed to our detailed results, we note that some work has
already appeared on meson-meson scattering in PsV systems.
Numerical results for  
many light S-wave PsV meson channels
were previously reported
by Swanson \cite{S_annphys} using a similar quark model 
approach that incorporated
OGE spin-spin and
linear confinement forces.
Theoretical results 
for PsV scattering 
($\pi\rho$ in particular)
in a meson exchange model
were published by 
Janssen {\it et al.}
\cite{Janssen}
and
B\"ockmann {\it et al.}
\cite{Boeckmann},
assuming
$\pi$, vector and $a_1$ exchange.
Since the 
$\rho\,\pi\pi$,
$a_1\rho\,\pi$ and 
$\rho\,\omega\pi$ 
vertex strengths
are relatively well established, it was possible to 
evaluate these scattering amplitudes  
numerically. These papers did not consider the exotic I=2
channel, so a direct comparison with our quark model PsV
results is not possible at present. 

\section{Meson-Meson T-matrix}

\subsection{General T-matrix formula}

We approximate the full hadron-hadron scattering
amplitude by a single (Born-order) matrix element
of the quark-quark interaction 
Hamiltonian $H_I$. Since $H_I$ is $T^a T^a$ in color,
one must then have quark line rearrangement
to have a nonvanishing overlap with two color-singlet mesons 
in the final state.
In $(q\bar q)-(q\bar q)$ scattering there are four independent
Born-order diagrams, which we label
according to which pair of
constituents interacted; these are 
``transfer$_1$" (T1), 
``transfer$_2$" (T2), 
``capture$_1$" (C1) 
and
``capture$_2$" (C2), 
which are shown in Fig.1. 
In the special case of identical quarks 
{\it and } identical antiquarks, which is relevant here,
there is a second set of four ``symmetrizing" diagrams
T1$_{symm} \dots$
C2$_{symm}$, which are identical to
T1$\dots$C2 except that the quark lines are interchanged rather
than the antiquark lines.

The hadron-hadron T-matrix element $T_{fi}$ for each diagram 
can conveniently be written
as an overlap integral of the meson wavefunctions times the 
underlying quark
$T_{fi}$. These overlap integrals (specializing 
Ref.\cite{BB} to the 
case of equal quark and antiquark masses) are 

\begin{eqnarray}
& &
T_{fi}^{\rm (T1)}(AB\to CD)  = 
\nonumber
\\
& &
\hskip 1cm
\int\! \! \! \int  d^3q \, d^3p \
\Phi_C^*(2\vec p + \vec q -  \vec C\, ) \;
\Phi_D^*(2\vec p - \vec q -2\vec A -  \vec C \, )  
\nonumber
\\
& &
\hskip 1cm
T_{fi}(\vec q, \vec p, \vec p - \vec A - \vec C   \, ) \  
\Phi_A(2\vec p - \vec q -  \vec A\, ) \;
\Phi_B(2\vec p + \vec q -  \vec A -2\vec C\, ) \ ,
\label{T1def}
\\
& &
\nonumber
\\
& &
T_{fi}^{\rm (T2)}(AB\to CD)  = 
\nonumber
\\
& &
\hskip 1cm
\int\! \! \! \int  d^3q \, d^3p \
\Phi_C^*(-2\vec p + \vec q +2\vec A - \vec C\, ) \;
\Phi_D^*(-2\vec p - \vec q -  \vec C \, )  
\nonumber
\\
& &
\hskip 1cm
T_{fi}(\vec q, \vec p, \vec p -\vec A + \vec C \, ) \  
\Phi_A(-2\vec p + \vec q + \vec A\, ) \;
\Phi_B(-2\vec p - \vec q + \vec A -2\vec C\, ) \  .
\label{T2def}
\\
& &
\nonumber
\\
& &
T_{fi}^{\rm (C1)}(AB\to CD)  = 
\nonumber
\\
& &
\hskip 1cm
\int\! \! \! \int  d^3 q \, d^3p \
\Phi_C^*(2\vec p + \vec q -  \vec C\, ) \;
\Phi_D^*(2\vec p - \vec q -2\vec A - \vec C \, )  
\nonumber
\\
& &
\hskip 1cm
T_{fi}(\vec q, \vec p, - \vec p + \vec C \, ) \  
\Phi_A(2\vec p - \vec q -  \vec A\, ) \;
\Phi_B(2\vec p - \vec q -  \vec A -2\vec C\, ) \ ,
\label{C1def}
\\
& &
\nonumber
\\
& &
T_{fi}^{\rm (C2)}(AB\to CD)  = 
\nonumber
\\
& &
\hskip 1cm
\int\! \! \! \int  d^3q \, d^3p \
\Phi_C^*(-2\vec p + \vec q +2\vec A - \vec C\, ) \;
\Phi_D^*(-2\vec p - \vec q -  \vec C \, )  
\nonumber
\\
& &
\hskip 1cm
T_{fi}(\vec q, \vec p, -\vec p -\vec C  \, ) \    
\Phi_A(-2\vec p + \vec q +  \vec A\, ) \;
\Phi_B(-2\vec p + \vec q +  \vec A -2\vec C\, ) \ .
\label{C2def}
\end{eqnarray}
The quark $T_{fi}$ has momentum arguments
$T_{fi}(\vec q, \vec p_1, \vec p_2)$, 
which are defined in Fig.2. 
In this paper we will evaluate these overlap integrals with 
standard Gaussian quark model wavefunctions and the quark $T_{fi}$
for 
the complete set of OGE color Coulomb, linear scalar, OGE spin-spin,
OGE and linear scalar confinement spin-orbit and OGE tensor
interactions. These interactions are given in App.A.

\subsection{PsPs Scattering}

\subsubsection{I=2 $\pi\pi$ T-matrix}

We specialize the general problem of 
PsPs scattering without $q\bar q$ annihilation 
to I=2 $\pi\pi$ because many experiments have
published phase shift analyses of this channel.
The other $\pi\pi$ channels have large $s$-channel 
$q\bar q$ annihilation contributions. 
The full 
I=2 $\pi\pi$ 
Born-order
T-matrix element is
determined by adding the individual contributions of App.B,
with PsPs spin matrix elements given in App.C, part C2. There
are also flavor and color factors for each diagram and an overall
``signature" phase of $(-1)$, and a second set of
``symmetrizing" diagrams for identical quarks and identical antiquarks,
as discussed in detail in Ref.\cite{pipi}. On summing 
these contributions we find

\begin{eqnarray}
&
T_{fi}^{I=2\; \pi\pi} 
=
&
+{\pi \alpha_s\over m^2}
\Bigg(
\; {2^3 \over 3^2}
\bigg(
e^{- Q_+^{\, 2} / 8\beta^2}
+
e^{- Q_-^{\, 2} / 8\beta^2}
\bigg)
+
{2^7 \over 3^{7/2}}\;
e^{- \vec A^{\, 2} / 3\beta^2}
\Bigg)
\nonumber\\
&
&
+ {\pi \alpha_s\over \beta^2}
\Bigg(
-
\ {2^4 \over 3^2}
\bigg(
{\rm f}_{{1\over 2},{3\over 2}}(\vec Q_+^{\, 2} /8\beta^2) 
+{\rm f}_{{1\over 2},{3\over 2}}(\vec Q_-^{\, 2} /8\beta^2) 
\bigg)
+{2^6 \over 3^{5/2}}\,
{\rm f}_{{1\over 2},{3\over 2}}(\vec A^{\, 2} /6\beta^2) 
\Bigg) \;
e^{- \vec A^{\, 2} / 2\beta^2}
\nonumber\\
&
&
+{\pi b\over \beta^4}
\Bigg(
\; {2^3\over 3}
\bigg(
{\rm f}_{-{1\over 2},{3\over 2}}(\vec Q_+^{\, 2} /8\beta^2) 
+
{\rm f}_{-{1\over 2},{3\over 2}}(\vec Q_-^{\, 2} /8\beta^2) 
\bigg)
-{2^3\over 3^{1/2}}\,
{\rm f}_{-{1\over 2},{3\over 2}}(\vec A^{\, 2} /6\beta^2) 
\Bigg) \;
e^{- \vec A^{\, 2} / 2\beta^2}
\;
\label{pipiT} 
\end{eqnarray}
where 
${\rm f}_{a,c}(x)$
is an abbreviation for
the confluent hypergeometric (Kummer) function
${}_1{\rm F}_1(a;c;x)$.

The three separate expressions above are the 
OGE spin-spin, 
color Coulomb and
linear confinement
contributions 
respectively. 
The 
$Q_\pm$ 
terms come from the transfer diagrams,
and the remaining, isotropic, terms 
come from the capture diagrams.
The spin matrix elements of the
spin-orbit and tensor terms vanish identically in 
the
PsPs channel.

Since
$\vec Q_\pm  = \vec C \pm \vec A$
and
$|\vec A | = | \vec C | $,
one can equivalently write this amplitude 
as a function of the
CM momentum and scattering angle using
$\vec Q_\pm^{\, 2} = 2 \vec A^{\, 2}(1\pm \mu)$, where
$\mu = \cos(\theta_{AC})$. 
The Bose symmetry required for this $\pi\pi$ scattering amplitude is evident.

\subsubsection{I=2 $\pi\pi$ Phase Shifts}

We may derive the elastic Born-order I=2 $\pi\pi$ phase shifts from
Eq.(\ref{pipiT}), 
using the relation between phase shifts and the T-matrix given
in App.D, especially Eq.(D17), and the integrals in App.G. 
The result we find for the 
S-wave is
\begin{equation}
\delta^{I=2\ \pi\pi}_0 
=
\cases{
kE_{\pi}  
{\alpha_s\over m^2}\;
\bigg(
-{1\over 3^2}\;
{1\over x} 
\Big(
1-e^{-2x}
\Big) 
-{2^4\over 3^{7/2}}\,  e^{-4x/3}
\bigg) 
&{\rm OGE spin-spin}
\cr 
kE_{\pi} 
{\alpha_s \over \beta^2}\; 
\bigg(
-{2\over 3^2}
{1\over x}
\Big(
{\rm f}_{1,{1\over 2}}(-2x) 
-e^{-2x}
\Big)
-{2^3 \over 3^{5/2}}\; 
{\rm f}_{1,{3\over 2}}(-2x/3)\, e^{-4x/3} 
\bigg)
\, 
\;
&{\rm OGE color Cou.} \cr 
kE_{\pi} 
{b\over \beta^4}\; 
\bigg(
{1\over 3^2}
{1\over x}
\Big(
{\rm f}_{2,{1\over 2}}(-2x) 
-e^{-2x}
\Big)
+
{1\over 3^{1/2}}\;
{\rm f}_{2,{3\over 2}}(-2x/3)\, e^{-4x/3} 
\bigg)\ . 
\,
\;
&{\rm linear conft.} 
}
\label{pipiSwave} 
\end{equation}
where we have introduced 
$x = {\vec A\,}^2 / 4\beta^2$. 
The total Born-order S-wave phase shift is the sum of these
three contributions. 

This
S-wave phase shift 
is shown in Fig.3
with our standard quark model parameter set
$\alpha_s=0.6$,
$\beta = 0.4$~GeV,
$m = 0.33$~GeV and
$b = 0.18$~GeV$^2$. We also use $M_\pi = 0.138$~GeV throughout.
This confirms that the color Coulomb and linear 
confinement interactions make relatively small contributions to the
I=2 $\pi\pi$ S-wave at moderate energies.
The weakly repulsive linear confining interaction 
in I=2 $\pi\pi$ 
near 
threshold 
was previously found numerically by Swanson\cite{S_annphys}.

One might be concerned about the approximation of using SHO wavefunctions,
especially at higher energy scales where there should be strong 
short-distance components in the pion wavefunction due to the attractive
spin-spin hyperfine interaction. To test the 
sensitivity to SHO wavefunctions 
we evaluated the I=2 $\pi\pi$ scattering amplitudes
and phase shifts numerically using Coulomb plus linear plus
hyperfine $q\bar q$ wavefunctions and Monte Carlo integration of
the real space integrals corresponding to the T-matrix
integrals 
(\ref{T1def}-\ref{C2def}). 
As usual this requires a ``smearing'' of the contact
hyperfine term,
$ \delta(\vec x) \to 
e^{-\sigma^2 r^2}
/\pi^{3/4} \sigma^{3/2}$,
to allow solution of the Schr\"odinger equation with an attractive 
delta-function interaction. 
In the literature the inverse smearing length is
typically 
taken to be $\sigma \approx 1$~GeV. 
(A calculation of I=2 $\pi\pi$ scattering with this interaction
and $\sigma=0.7$~GeV
was reported previously by Swanson
\cite{S_annphys}.) 
With our standard light-quark parameter set
$\alpha_s=0.6$, 
$b=0.18$~GeV$^2$ and
$m=0.33$~GeV,
we found 
that fitting 
the $M_\rho - M_\pi$ splitting
required a value of
$\sigma = 0.86$~GeV. 
To illustrate the dependence of the scattering amplitude on this
parameter, in Fig.4 we show the 
I=2 $\pi\pi$
S-wave 
that follows from our standard quark model set $(\alpha_s, b, m)$,
with 
$\sigma =
0.7, 0.8$ and $0.9$~GeV. 
Clearly the predicted phase shift is rather similar to the SHO result
of Fig.3,
although the effect of 
short-distance peaking in the $\pi$ wavefunction is 
evident above $M_{\pi\pi}\approx 1$~GeV. 

We also show most of the higher statistics experimental results
for the I=2 $\pi\pi$
S-wave  phase shift in Figs.3 and 4. The references shown are 
Colton {\it et al.}\cite{Colton},
Durusoy {\it et al.}\cite{Durusoy} 
(OPE extrapolation, solid; OPE + DP form factor, open, slightly displaced in
$x$ for visibiity),
Hoogland {\it et al.}\cite{Hoogland}
(extrapolation B), 
and Losty {\it et al.}\cite{Losty}.
Prukop {\it et al.}\cite{Prukop} found a wide range of results
from three different off-shell extrapolations, so we simply quote their
fitted scattering length below.

Clearly 
there is already reasonable agreement with the experimental S-wave phase shift
at lower energies without fitting the quark model parameters. 
The model predicts a rather dramatic extremum in this phase shift near
$M_{\pi\pi}=1.5$~GeV, which is unfortunately beyond the limiting
invariant mass of most of the experimental studies.
There are some 
measurements of this phase shift at higher invariant mass 
with lower accuracy due to
Durusoy {\it et al.}\cite{Durusoy}, which are
also shown in the figure.
The  
Durusoy {\it et al.} results support
our predicted extremum near $M_{\pi\pi}=1.5$~GeV; indeed, 
their phase shift 
above 
$M_{\pi\pi}=1.5$~GeV 
appears to fall even more rapidly than 
we predict. 

We have investigated optimal parameter fits of the S-wave phase shift formula
Eq.(6) to the data,
but we find that these are
rather unstable because the color Coulomb and linear confinement 
contributions are
small and are qualitatively similar functions. 
In any case the Durusoy {\it et al.} data and Fig.4 show that the 
hyperfine smearing distance $\sigma$ is an important parameter, and this
will not be well determined until accurate phase shift data
becomes available at higher invariant mass. 
An accurate measurement of I=2 $\pi\pi$ scattering
amplitudes near and above 
$M_{\pi\pi}=1.5$~GeV 
would 
clearly be very useful as a test of 
this and other models of meson-meson scattering.

\subsubsection{I=2 $\pi\pi$ Scattering Lengths}

The I=2 $\pi\pi$ scattering 
length is defined by $a_0^{I=2} = \lim_{k_\pi\to 0} \delta_0^{I=2\; \pi\pi} / k_\pi$.
The results we find from Eq.(6) are
\begin{equation}
a_0^{I=2} 
=
\cases{
-{2\over 3^2}\; 
\bigg(
1 + {2^3\over 3^{3/2}}
\bigg) \; 
{\alpha_s \over m^2}\;
M_{\pi}
&{$S\cdot S$}  
\cr
-{2^2\over 3^2}\; 
\bigg(
{2\over 3^{1/2} }-1
\bigg) \;
{\alpha_s  \over \beta^{\, 2}}\;
M_{\pi}
&{Cou.} 
\cr 
-{2\over 3}\; 
\bigg(
1-{3^{1/2}\over 2}\;
\bigg) \; 
{b \over \beta^{\, 4}}\;
M_{\pi}
&{lin.}  
}
\end{equation}
and their numerical values with our standard quark model parameters set
are
\begin{equation}
a_0^{I=2} 
=
\cases{
-0.085 
\ {\rm [fm]}
&{$S\cdot S$}  
\cr 
-0.007
\ {\rm [fm]}
&{Cou.} 
\cr 
-0.017
\ {\rm [fm]}
&{lin.}  
\cr 
-0.109
\ {\rm [fm]}
&{\rm total}.
}
\end{equation}
The Coulomb and linear contributions were independently checked by
Monte Carlo integration of the corresponding real-space overlap integrals.
The relative sizes of these numerical contributions 
{\it a posteriori} 
justify the approximation of 
neglecting the color Coulomb and linear terms
in 
I=2 $\pi\pi$ scattering. 

The I=2 $\pi\pi$ scattering length has been calculated previously
using many
other theoretical approaches. A summary of some of these predictions
is given below.
(We use a current value of
$f_\pi = 93$~MeV in Weinberg's PCAC 
formula $a_0^{I=2} = -M_\pi / 16\pi {f_\pi}^2$.)

\begin{equation}
a_0^{I=2}\bigg|_{thy.} 
=
\cases{
-0.053(7)
\ {\rm [fm]}
&{LGT \cite{LGT_pipi}} 
\cr 
-0.052 
\ {\rm [fm]}
&{meson exchange \cite{mesex,FS}}
\cr 
-0.053 
\ {\rm [fm]}
&{Roy Eqs.\cite{Ana}}  
\cr 
-0.063  
\ {\rm [fm]}
&{PCAC \cite{Wein}}.  
}  
\end{equation}

Although $f_\pi$ and other effective lagrangian parameters are
normally taken from experiment, these parameters are of course
calculable from quark-gluon forces. As an example, our
result for $a_0^{I=2}$ yields the following expression for $f_\pi$,
\begin{equation}
{1\over {f_\pi}^2 } =
{2^5\over 3^2}\; 
\bigg(
1 + {2^3\over 3^{3/2}}
\bigg) \; 
{\pi \alpha_s \over m^2}\;
+{2^6\over 3^2}\; 
\bigg(
{2\over 3^{1/2} }-1
\bigg) \;
{\pi \alpha_s \over \beta^{\, 2}}\;
+{2^5\over 3}\; 
\bigg(
1-{3^{1/2}\over 2}\;
\bigg) \; 
{\pi b \over \beta^{\, 4}} \ .
\end{equation}
The dominant contribution comes from the 
$O(\alpha_s / m^2 )$ OGE $S\cdot S$ term.

Experimental determinations of the scattering length have yielded
results which are larger than theoretical expectations; 
\begin{equation}
a_0^{I=2}\bigg|_{expt.} 
=
\cases{
-0.13(2) 
\ {\rm [fm]}
&{Losty {\it et al.}\cite{Losty}}  
\cr 
-0.24(2), -0.22{+0.03\atop -0.04}
\ {\rm [fm]}
&{Prukop {\it et al.} \cite{Prukop}}  \ .
}
\end{equation}
We speculate that this discrepancy is due to the use of a simple 
$\delta = k_\pi a + O({k_\pi}^3)$
effective range formula in extrapolation. The
difficulty of extrapolating experimental phase shifts to threshold
has been stressed by Morgan and Pennington \cite{MP,Penn}.
We advocate the use of a ``generalized specific heat plot"
of low energy phase shifts for this purpose \cite{KN}. 
This plot takes into account the threshold behavior seen in Eq.(6),
\begin{equation}
\delta_0^{I=2 \pi\pi} =
k_\pi E_\pi \; f(\alpha_s, m, b, \beta, x)
\end{equation}
where $f$ is a relatively slowly varying function of 
$x = {k_\pi}^2/ 4 \beta^2$. Thus the threshold behavior is approximately
proportional to $k_\pi E_\pi$ rather than just $k_\pi$, and since $M_\pi$ 
is quite small this leads to rapid variation 
near threshold and makes the linear-$k_\pi$ extrapolation inaccurate.
We suggest removal of all this dependence by displaying 
$\delta_0 / (k_\pi E_\pi / M_\pi ) $ versus ${k_\pi}^2$. 
The intercept in this plot is the scattering length,
and the slope at intercept implies the effective range. 

This generalized specific heat plot is shown in Fig.5 
for I=2 $\pi\pi$ scattering.
An extrapolation of the moderate-$k_\pi$ data
can now be seen to be much closer to
the theoretical 
scattering lengths.
The small-$k_\pi$
dependence of  $\delta_0/k_\pi$ 
was calculated by Donoghue \cite{Don} in a chiral effective
lagrangian, which gave the Weinberg result at ${k_\pi}=0$ and an
$O(k_\pi^2)$ correction factor of
$(1 + {k_\pi}^2/2 {m_\pi}^2) $. This is just the correction due to
an overall factor of $E_\pi$, so this predicts a zero
slope in ${k_\pi}^2$ 
for $\delta_0 / (k_\pi E_\pi / M_\pi ) $  
at threshold. 

The J\"ulich meson-exchange model \cite{mesex}, 
which is dominated by $t$-channel
$\rho$ exchange in this process, also predicts rapid variation
in $\delta_0/k_\pi$ near threshold. The prediction of this model
for $\delta_0/(k_\pi E_\pi / M_\pi )$ \cite{FS}, 
shown in Fig.5, is rather similar
to our quark model result. 

\subsubsection{I=2 $\pi\pi$ Equivalent Potentials}

Low energy ``phase shift equivalent'' Gaussian 
I=2 $\pi\pi$ potentials, derived using the method
of Mott and Massey \cite{MM} as described in App.E,
are given below. We quote separate Gaussians for the transfer and
capture contributions from each of the three interactions, spin-spin
contact, color Coulomb and linear hyperfine.  However their predicted
phase shift decays more slowly at large momentum, probably due to the
use of power law form factors in their vertices.
\begin{equation}
V_{\pi\pi}(r) 
=
\cases{
+{2^{9/2}\over 3^2{\pi}^{1/2}}\;  
{\alpha_s \beta^3\over m^2}\;
e^{-2\beta^2 r^2}
&{$S\cdot S$\ (transfer)} \cr 
+{2^{9/2}\over 3^2{\pi}^{1/2}}\;  
{\alpha_s \beta^3\over m^2}\;
e^{-{3\over 2}\beta^2 r^2}
&{$S\cdot S$\ (capture)} 
\cr 
-{2^{11/2}\over 3^{1/2} 5^{3/2} {\pi}^{1/2}}\;  
{\scriptstyle \alpha_s \beta}\;
e^{-{6\over 5}\beta^2 r^2}
&{Cou.\ (transfer)} \cr 
+{2^{1/2}3^{1/2} \over {\pi}^{1/2}}\;  
{\scriptstyle \alpha_s \beta}\;
e^{-{9\over 8}\beta^2 r^2}
&{Cou.\ (capture)} \cr 
+{2^{9/2} 3^{1/2}\over 7^{3/2} {\pi}^{1/2}}\;  
{b\over \beta}\;
e^{-{6\over 7}\beta^2 r^2}
&{lin.\ (transfer)} \cr 
-{2^{1/2} 3^{5/2}\over 5^{3/2} {\pi}^{1/2}}\;  
{b\over \beta}\;
e^{-{9\over 10}\beta^2 r^2}
&{lin.\ (capture)}. 
}
\end{equation}

In
Ref.\cite{pipi} 
we derived I=2 $\pi\pi$ potentials 
for the spin-spin contact interaction 
using the ``locality expansion" method of
Ref.\cite{BG}. This gave an identical result for the 
spin-spin transfer potential, because this amplitude 
(before Bose symmetrization) is a
function of $t$ only. However for the 
capture diagrams the Mott-Massey approach used here 
gives a different potential, since it is
constrained to reproduce the $O(k^3)$ 
series expansion of the phase shift in Eq.(6), but the
local approximation is not. The two capture potentials 
reproduce the 
scattering length, but the local approximation gives an incorrect
effective range.

The low energy Mott-Massey I=2 $\pi\pi$ potential is shown in Fig.6 
for our standard quark model 
parameters $\alpha_s=0.6$, 
$b=0.18$~GeV$^2$ and 
$m=0.33$~GeV. The spin-spin hyperfine contribution is dominant 
over the range shown.

\subsubsection{I=2 $\pi\pi$ Phase Shifts with L $>$ 0}

The higher 
partial waves 
(L $\geq  2$)
may be evaluated similarly.
According to Eq.(\ref{pipiT}), these receive contributions only from
the transfer
diagrams. The Born-order D-wave phase shift with SHO wavefunctions 
is given by

\begin{equation}
\delta^{I=2\ \pi\pi}_2 
=
\cases{
kE_{\pi}  
{\alpha_s\over m^2}\;
\bigg(
-{1\over 3^2}\; 
{1-e^{-2x}\over x} 
+{1\over 3}\; 
{1+e^{-2x}\over x^2} 
-{1\over 3}\; 
{1-e^{-2x}\over x^3} 
\bigg) 
& $S\cdot S$
\cr
kE_{\pi} 
{\alpha_s \over \beta^2}\; 
\bigg(
-{2\over 3^2}\;
{  {\rm f}_{1,{1\over 2}}(-2x) -e^{-2x}  \over  x  }
+{2\over 3^2}\;
{  {\rm f}_{1,-{1\over 2}}(-2x) +e^{-2x}  \over  x^2  }
-{2\over 3\cdot 5 }\;
{  {\rm f}_{1,-{3\over 2}}(-2x) -e^{-2x}  \over  x^3  }
\bigg)
&{\rm Cou.} 
\cr 
kE_{\pi} 
{b \over \beta^4}\; 
\bigg(
{1\over 3}\;
{  {\rm f}_{2,{1\over 2}}(-2x) -e^{-2x}  \over  x  }
-{1\over 5}\;
{  {\rm f}_{2,-{1\over 2}}(-2x) +e^{-2x}  \over  x^2  }
+{3\over 5\cdot 7 }\;
{  {\rm f}_{2,-{3\over 2}}(-2x) -e^{-2x}  \over  x^3  }
\bigg) \ .
&{\rm lin.}  
}
\label{pipiDwave} 
\end{equation}
These three expressions are numerically rather small, and their phases 
are such that they approximately cancel;
at $M_{\pi\pi}=1.5$~GeV they are respectively
$-0.8^o$, 
$+0.2^o$ and
$+0.4^o$. 
To see this more clearly, the leading
$O({k_\pi}^5)$ behavior predicted by Eq.(14) is
\begin{equation}
\lim_{k_\pi \to 0 } \delta^{I=2\ \pi\pi}_2  /  {k_\pi}^5 
=
{1\over 2^3\, 3^3\, 5^2}\; 
\bigg(
-5\,
{\alpha_s\beta^2 \over m^2 }\;
+2\, \alpha_s \; 
+3\, {b \over \beta^2}\; 
\bigg) \; {M_\pi\over \beta^6 }
\ ,
\end{equation}
and the three dimensionless combinations
$\alpha_s\beta^2 /  m^2$,
$\alpha_s$ and
$b/ \beta^2$ 
are comparable in size.
We have also evaluated this 
D-wave phase shift 
using 
Coulomb plus linear plus hyperfine wavefunctions. The result
is shown in Fig.4, and is
numerically similar to the SHO D-wave, 
Eq.(14). 

In comparison the experimental D-waves reported by
Durusoy {\it et al.} \cite{Durusoy} and
Hoogland {\it et al.} \cite{Hoogland}
are $\approx -3^o$ at $M_{\pi\pi}=1.5$~GeV (see Fig.4). 
(Losty {\it et al} \cite{Losty} report a rather larger but
inconsistent low-energy D-wave.)
This is clearly larger than our prediction, although the
rather slow variation of the Durusoy {\it et al.} and Hoogland {\it et al.}
D-waves with  $M_{\pi\pi}$ may indicate a problem with the measurements;
the expected threshold behavior of ${k_\pi}^5$ is much more rapid 
than the 
observed energy dependence. Unfortunately the dispersion relations 
represented by the Roy equations have technical difficulties with
determining D- and higher waves \cite{Ana}.
They do however lead to predictions of a {\it positive}
D-wave close to threshold,
which is not evident in the data.
The D-wave may well have important meson exchange contributions,
since this type of
model can accommodate the reported experimental phase shift \cite{SK}.

\subsection{PsV Scattering}

\subsubsection{I=2 $\pi\rho$ T-matrix}

For simplicitly we will initially quote results only for 
I=2 $\pi\rho$.
The other isospin channels are simply related by flavor
factors, which we will discuss subsequently. 
We assume identical spatial wavefunctions,
so only the $\rho$ spin degree of freedom and difference in phase space 
distinguish this 
case from
$\pi\pi$.
Summing the individual contributions in
App.B with the appropriate flavor and color factors
and the $(-1)$ signature phase, and using the PsV spin matrix elements of 
App.C part C3, we find for the I=2 $\pi\rho$ T-matrix 

\begin{displaymath}
T_{fi}^{I=2\; \pi\rho} =
\hfill
\end{displaymath}
\begin{displaymath}
+
{\pi \alpha_s \over m^2 }
\Bigg(
+{2^3 \over 3^3}
\;
\bigg(
3\, e^{ - \vec Q_-^{\, 2}/8\beta^2 }
\; 
-
e^{ - \vec Q_+^{\, 2}/8\beta^2 }
\bigg)
+{2^7 \over 3^{9/2}}
\; e^{ - \vec A^{\, 2}/3\beta^2 }
\Bigg)
\hfill
\end{displaymath}
\begin{displaymath}
+
{\pi \alpha_s\over \beta^2}
\Bigg(
-{2^4 \over 3^2}
\bigg(
\;
{\rm f}_{{1\over 2},{3\over 2}}(\vec Q_-^2 /8\beta^2) 
+
{\rm f}_{{1\over 2},{3\over 2}}(\vec Q_+^2 /8\beta^2) 
\bigg)
+
{2^6 \over 3^{5/2}}
\;
{\rm f}_{{1\over 2},{3\over 2}}(\vec A^{\, 2} /6\beta^2) 
\Bigg)
\; e^{ - \vec A^{\, 2}/2\beta^2 }
\hfill
\end{displaymath}
\begin{displaymath}
+
{\pi b\over \beta^4}
\Bigg(
+{2^3 \over 3}
\bigg(
{\rm f}_{-{1\over 2},{3\over 2}}(\vec Q_-^2 /8\beta^2) 
+
{\rm f}_{-{1\over 2},{3\over 2}}(\vec Q_+^2 /8\beta^2) 
\bigg)
-{2^3 \over 3^{1/2}}
\;
{\rm f}_{-{1\over 2},{3\over 2}}(\vec A^{\, 2} /6\beta^2) 
\Bigg)
\; e^{ - \vec A^{\, 2}/2\beta^2 }
\hfill
\end{displaymath}
\begin{displaymath}
+
{\pi \alpha_s\over m^2 \beta^2}
\Bigg(
-{2^2 \over 3^{2}}
\;
{\rm f}_{{3\over 2},{5\over 2}}(\vec Q_+^2 /8\beta^2) 
-{2^4 \over 3^{9/2}}
\;
{\rm f}_{{3\over 2},{5\over 2}}(\vec A^{\, 2} /6\beta^2) 
\Bigg)
\; e^{ - \vec A^{\, 2}/2\beta^2 }
\bigg[
\; \vec S_\rho \cdot i( \vec A \times \vec C \; )
\bigg]
\hfill
\end{displaymath}
\begin{displaymath}
+
{\pi b\over m^2 \beta^4}
\Bigg(
+{2 \over 3^{2}}
\;
{\rm f}_{{1\over 2},{5\over 2}}(\vec Q_+^2 /8\beta^2) 
-{2 \over 3^{5/2}}
\;
{\rm f}_{{1\over 2},{5\over 2}}(\vec A^{\, 2} /6\beta^2) 
\Bigg)
\; e^{ - \vec A^{\, 2}/2\beta^2 }
\bigg[
\; \vec S_\rho \cdot i( \vec A \times \vec C \; )
\bigg]
\hfill
\end{displaymath}
\begin{displaymath}
+
{\pi \alpha_s\over m^2 \beta^2}
\Bigg(
+
{2 \over 3^2\cdot 5}
\;
{\rm f}_{{5\over 2},{7\over 2}}(\vec Q_-^{\, 2} /8\beta^2) 
\Bigg)
\; e^{ - \vec A^{\, 2}/2\beta^2 }
\; \bigg[ \vec S_\rho \cdot \vec Q_-\;\vec S_\rho \cdot \vec Q_-   
- {2\over 3}\vec Q_-^{\, 2} \bigg] 
\hfill
\end{displaymath}
\begin{displaymath}
+
{\pi \alpha_s\over m^2 \beta^2}
\Bigg(
+
{2^5 \over 3^{9/2}\cdot 5}
\;
{\rm f}_{{5\over 2},{7\over 2}}(\vec A^{\, 2} /6\beta^2) 
\Bigg)
\; e^{ - \vec A^{\, 2}/2\beta^2 }
\Bigg[
\; \bigg[ \vec S_\rho \cdot \vec A\;\vec S_\rho \cdot \vec A   
- {2\over 3}\vec A^{\, 2} \bigg]
+\bigg[ \vec S_\rho \cdot \vec C\;\vec S_\rho \cdot \vec C   
- {2\over 3}\vec C^{\, 2} \bigg]
\Bigg]
\hfill
\ .
\end{displaymath}
\begin{equation}
\label{pirhoT}
\end{equation}
The individual contributions in this result are respectively
OGE spin-spin; 
OGE color Coulomb; 
linear confinement;
OGE spin-orbit; 
linear scalar confinement spin-orbit;
OGE tensor (transfer diagrams); 
and 
OGE tensor (capture diagrams). 
In all these we list
transfer followed by capture contributions.
$\vec S_\rho$ is the $\rho$ meson spin vector,
$\vec A$ and $\vec C$ are the initial and final $\pi$ momenta,
$\vec B = - \vec A$ and 
$\vec D = - \vec C$ are the initial and final $\rho$ momenta, 
and 
$\vec Q_\pm = \vec C \pm \vec A$ as in $\pi\pi$.
Since this result was derived in the CM frame,
$|\vec A | = |\vec B | = |\vec C | = |\vec D |$.
This $T_{fi}$ evidently describes
$\pi\rho$ 
spin-orbit and tensor interactions, 
in addition to spin-independent scattering.
It is interesting that there is a one-to-one mapping
between the 
quark-quark 
spin-orbit and tensor 
interactions
and these 
$\pi\rho$ 
spin-orbit and tensor 
terms. This simple result need not be true in general; 
a given spin-dependent interaction at the quark level 
may give rise to a different type of 
hadron-hadron interaction. As an example, a mapping of a 
tensor nucleon-nucleon force
into a nucleon-nucleus spin-orbit interaction
was discussed by 
Stancu, Brink and Flocard.\cite{SBF}

To evaluate phase shifts and inelasticities it is convenient to 
calculate the
matrix element of our $\pi\rho$ T-matrix
Eq.(\ref{pirhoT}) between general $|jls\rangle$ states, which gives the
reduced matrix element

\begin{displaymath}
T^j_{l'l} \equiv  
\langle jm, l's| T | jm,ls\rangle  = 
\hskip 3cm
\end{displaymath}
\begin{equation}
\sum_{ \mu \mu'\atop s_z s_z'} 
\langle         
jm
\vert 
l' \mu', 1 s_z' 
\rangle 
\langle 
jm
\vert 
l \mu, 1 s_z 
\rangle 
\int\!\!\! \int  d\Omega' \, d \Omega \;
Y^*_{l' \mu'}(\Omega') \; 
\langle 
1 s_z' 
\vert 
T_{fi} (\Omega',\Omega)  
\vert 
1 s_z 
\rangle \;
Y_{l \mu}(\Omega) \ ,
\end{equation}
as discussed in App.D. This is a straightforward exercise, although
integrals of special functions and a careful angular analysis
of the spin-orbit and tensor terms are required; the details are
discussed in Apps.G and H. This matrix element is diagonal in $l$
except for the tensor interaction, which has both diagonal and off-diagonal
(transfer) and fully off-diagonal (capture) contributions.
The $l$-diagonal results, again showing transfer diagram contributions
followed by capture, are

\begin{eqnarray}
&&T^j_{ll}|_{S\cdot S}  =  
{\pi^2 \alpha_s \over m^2 }\; 
\left(  
(1 + \delta_{l,odd} ) \; 
{2^6 \over 3^3}\; 
i_l(x)\; 
e^{-x}
+ \delta_{l,0}\; {2^9 \over 3^{9/2}}\;
e^{-4x/3}
\right)
\\
& &  \nonumber \\
&&T^j_{ll}|_{Cou.}  =  
{\pi^2 \alpha_s \over \beta^2}\; 
\left( 
- \delta_{l,even} \; 
{2^6 \over 3^2}\; 
{\cal F}^{(l)}_{{1\over 2},{3\over 2}}(x) 
+ \delta_{l,0}\; {2^8 \over 3^{5/2}}\;
f_{{1\over 2},{3\over 2}}(2x/3)  
\right)
e^{-2x}
\\
& &  \nonumber \\
&&T^j_{ll}|_{lin.} = 
{\pi^2 b \over \beta^4}\; 
\left( 
\delta_{l,even} \; 
{2^5 \over 3}\; 
{\cal F}^{(l)}_{-{1\over 2},{3\over 2}}(x) 
- \delta_{l,0}\; {2^5 \over 3^{1/2}}\;
f_{-{1\over 2},{3\over 2}}(2x/3) 
\right)
e^{-2x}
\\
& &  \nonumber \\
& &  \nonumber \\
&&T^j_{ll}|_{OGE\ L\cdot S\, }  =  
{\pi^2 \alpha_s \over m^2 }\; 
\langle \vec {\rm L} \cdot \vec {\rm S}\,  \rangle \;
\nonumber \\
& &
\hskip2cm 
\cdot
\left(  
-{2^5 \over 3^2}\; 
{
1
\over 
(2l+1)
}
x\;
\Big(
{\cal F}^{(l-1)}_{{3\over 2},{5\over 2}}(x) 
-
{\cal F}^{(l+1)}_{{3\over 2},{5\over 2}}(x) 
\Big)
- \delta_{l,1}\; {2^8 \over 3^{11/2}}\;
x\;
f_{{3\over 2},{5\over 2}}(2x/3) 
\right)
e^{-2x}
\\
& &  \nonumber \\
&&T^j_{ll}|_{lin.\ L\cdot S }  =  
{\pi^2 b \over m^2 \beta^2 }\; 
\langle \vec {\rm L} \cdot \vec {\rm S}\, \rangle \;
\nonumber \\
& &
\hskip2cm 
\cdot
\left(  
{2^4 \over 3^2}\; 
{1
\over 
(2l+1)
}
x\;
\Big(
{\cal F}^{(l-1)}_{{1\over 2},{5\over 2}}(x) 
-
{\cal F}^{(l+1)}_{{1\over 2},{5\over 2}}(x) 
\Big)
-
\delta_{l,1}\; 
{2^5 \over 3^{7/2}}\;
x\;
f_{{1\over 2},{5\over 2}}(2x/3) 
\right)
e^{-2x}
\\
& &  \nonumber \\
&&T^j_{ll}|^{\it transfer}_{OGE\ T}  =  
{\pi^2 \alpha_s \over m^2 }\; 
\langle T \rangle
\;
\nonumber \\
& &
\cdot
(-1)^{l+1}
{2^4 \over 3^3\cdot  5 }\;
x\; 
\left(  
{l\over (2l+1)}
{\cal F}^{(l-1)}_{{5\over 2},{7\over 2}}(x) 
+ 
{2l\over (2l+3)}
{\cal F}^{(l)}_{{5\over 2},{7\over 2}}(x) 
+ 
{l(2l-1)\over (2l+1)(2l+3)}
{\cal F}^{(l+1)}_{{5\over 2},{7\over 2}}(x) 
\right)
e^{-2x}
\end{eqnarray}
and the off-diagonal tensor matrix elements are
\vskip0.5cm
\begin{eqnarray}
&&T^j_{l'\ne l}|^{\it transfer}_{OGE\ T}  =  
{\pi^2 \alpha_s \over m^2 }\; 
\Big(
\delta_{l,j-1} \delta_{l',j+1} 
+
\delta_{l,j+1} \delta_{l',j-1} 
\Big) 
\nonumber \\
&&\hskip2cm \cdot
(-1)^{j+1} 
{2^4 \over 3^2\cdot 5}\;
{[ j(j+1)]^{1/2} \over  (2j+1)\hskip0.3cm  }\;
x
\bigg(
{\cal F}^{(j-1)}_{{5\over 2},{7\over 2}}(x) 
+2 {\cal F}^{(j)}_{{5\over 2},{7\over 2}}(x) 
+{\cal F}^{(j+1)}_{{5\over 2},{7\over 2}}(x) 
\bigg)\;
\; e^{-2x} 
\\
&&
\nonumber \\
&&T^j_{l'\ne l}|^{\it capture}_{OGE\ T}  =  
{\pi^2 \alpha_s \over m^2 }\; 
\delta_{j1} \;
\Big(
\delta_{l2} \delta_{l'0} 
+
\delta_{l0} \delta_{l'2} 
\Big) 
\;
{2^{19/2} \over 3^{11/2}\cdot 5}\;
x\; 
f_{{5\over 2},{7\over 2}}(2x/3)
\; e^{-2x} \ .
\end{eqnarray}
\vskip0.5cm
In these formulas $i_l(x)$ is a modified
spherical Bessel function, 
the tensor $\langle T \rangle$ matrix element between 
$|j,l,s=1\rangle$ $\pi\rho$ states is
\begin{equation}
\langle T \rangle = 
\cases{
1
& $j=l+1$
\cr 
-(2l+3)/l
& $j=l$
\cr 
(l+1)(2l+3)/l(2l-1)
& $j=l-1$ \ ,
}
\end{equation}
and the integral  
\begin{equation}
{\cal F}^{(l)}_{a,c}(x) 
\equiv \int_{-1}^1 d \mu \, {\rm P}_l(\mu)\; {}_1{\rm F}_1(a;c;x(1+\mu))
\end{equation}
is evaluated
in App.G.

\subsubsection{I=2 $\pi\rho$ S-wave Phase Shifts}

In S-wave to S-wave scattering the spin-orbit and tensor 
$\pi\rho$ 
T-matrix contributions vanish, and 
we are left with color Coulomb, linear and spin-spin contributions,
just as in the I=2 $\pi\pi$ case. 
The I=2 $\pi\rho$ 
S-wave phase shifts that result from these interactions, again
using Eq.(D17), are
\begin{equation}
\delta^{I=2\ \pi\rho}_0 
=
\cases{
{k E_{\pi} E_{\rho} \over \sqrt{s} } \;
{\alpha_s\over m^2}\;
\bigg(
-{2^2\over 3^3}\;
{1\over x}
\Big( 1 - e^{-2x}\Big) 
-
{2^6 \over 3^{9/2}}\;
e^{-4x/3}
\bigg)
&{\rm $S\cdot  S$}
\cr
{k E_{\pi} E_{\rho} \over \sqrt{s} } \;
{\alpha_s \over \beta^2}\; 
\bigg(
-{2^3\over 3^2}\;
{1\over x}
\Big(
{\rm f}_{1,{1\over 2}}(-2x) - e^{-2x}  
\Big)
-{2^5 \over 3^{5/2}}\; 
{\rm f}_{1,{3\over 2}}(-2x/3)\, e^{-4x/3} 
\bigg)
&{\rm Cou.} 
\cr 
{k E_{\pi} E_{\rho} \over \sqrt{s} } \;
{b\over \beta^4}\; 
\bigg(
{2^2\over 3^2}\;
{1\over x}
\Big(
{\rm f}_{2,{1\over 2}}(-2x)  - e^{-2x}  
\Big)
+{2^2 \over 3^{1/2}}\; 
{\rm f}_{2,{3\over 2}}(-2x/3) \, e^{-4x/3}
\bigg)
&{\rm lin.} 
}
\label{pirhoSwave} 
\end{equation}
where $\sqrt{s} = (E_\pi + E_\rho)$, and again
$x = {\vec A\,}^2 / 4\beta^2$. 
In Fig.6 we show these individual components and the total 
S-wave phase shift with our standard quark model parameter set and
meson masses (used throughout) of 
$M_\pi=0.138$~GeV
and
$M_\rho=0.77$~GeV.
The forces considered here 
evidently lead to strong repulsion in the 
I=2 $\pi\rho$ channel. 

\subsubsection{I=2 $\pi\rho$ Phase Shifts with $L>0$}

The spin-orbit and tensor terms in
Eqs.(21-23)
all contribute to 
$l>0$
$\pi\rho$ 
scattering, 
and there is also an odd-$l$,  
$j$-independent
term due to the OGE spin-spin 
interaction in Eq.(18), which is not symmetric under $\mu \to -\mu$.
The color Coulomb and linear confinement 
spin-independent terms, Eqs.(19,20), contribute only to 
even $l$.

Adding the various diagonal matrix elements of Eqs.(18-23) and
using Eq.(D17) gives phase shifts for each $^3$L$_{\rm J}$ partial wave. 
In Fig.8 we show results for all P-wave channels and for
J=L$\pm 1$ in D- and F-wave.
Note that there is a large, inverted spin-orbit force in the P-wave,
so the $^3$P$_0$ phase shift is widely separated from 
$^3$P$_2$, and has an even larger maximum phase shift than the S-wave. 
The higher-L channels show decreasing phase shifts with 
increasing L, as expected
for short-ranged quark-gluon forces.

The relative importance of the 
individual contributions to the spin-dependent force is
of considerable interest. In Fig.9 we show the various 
spin-dependent contributions to the
I=2 $^3$P$_2$ $\pi\rho$ phase shift. The largest contribution arises from
OGE spin-orbit, in particular from the transfer diagrams. The OGE 
and confinement spin-orbit 
capture diagrams give smaller contributions
of the same sign. Finally, the confinement spin-orbit 
transfer
diagrams have a sign opposite to all these and reduces the total spin-orbit
force somewhat. This dominance of the PsV 
spin-orbit by OGE is an interesting result,
especially since Mukhopadhyay and Pirner \cite{MP_KN} found the opposite
result in KN. In that system they concluded that 
confinement, not OGE, makes the largest contribution to the spin-orbit
force.
The OGE tensor
in I=2 $^3$P$_2$ $\pi\rho$
is weakly repulsive; it makes a much larger contribution to
$^3$P$_1$ and $^3$P$_0$, where the tensor matrix element is 
respectively $-5$ and $10$ times as large. The OGE 
tensor is evident in Fig.8, in the departure of the ratio
($^3$P$_2-{}^3$P$_1$):($^3$P$_1-{}^3$P$_0$) from the pure spin-orbit value
of 2:1 at higher energies.

There is also an off-diagonal coupling due to the OGE tensor terms,
given by Eqs.(24,25), but 
we have neglected this in calculating 
phase shifts because we find that it is numerically
a small effect. The largest coupling at low energies is 
${}^3$S$_1 \leftrightarrow {}^3$D$_1$, which leads to an inelasticity of only
$\eta_{SD} = 0.97$ 
by $M_{\pi\rho} = 3.0$~GeV   
(calculated using Eqs.(D18-D20)). 

\subsubsection{I=2 $\pi\rho$ P-wave Spin-Orbit Potentials}

We may determine low energy Gaussian equivalent 
$\pi\rho$ potentials from the phase shifts, as discussed in App.E. 
The most interesting potential phenomenologically is the spin-orbit
one, since the origin of the spin-orbit interaction in the NN system
is a long-standing and still poorly understood problem.
In particular we derived Gaussian potentials corresponding to the 
P-wave phase shifts due to the
OGE and linear scalar confinement spin-orbit interactions,
using Eq.(E4) of App.E. The results for the 
transfer 
and 
capture
contributions to these potentials are

\begin{equation}
V^{s.o.}_{\pi\rho}(r) 
=
\cases{
- { 2^{11/2}  5^{5/2} \over 3^2 7^{5/2} \pi^{1/2} }\;
{\alpha_s \beta^3 \over m^2 } \;
\langle \vec {\rm L} \cdot \vec {\rm S} \rangle 
\; e^{-{10\over 7}\beta^2 r^2 }
&{OGE\ (transfer)} 
\cr 
- {  5^{5/2} \over 3^{9/2} \pi^{1/2} }\;
{\alpha_s \beta^3 \over m^2 } \;
\langle \vec {\rm L} \cdot \vec {\rm S} \rangle 
\; e^{-{5\over 4}\beta^2 r^2 }
&{OGE\ (capture)} 
\cr 
+ { 2^{9/2}  5^{5/2} \over 3^7 \pi^{1/2} }\;
{b \beta \over m^2 } \;
\langle \vec {\rm L} \cdot \vec {\rm S} \rangle 
\; e^{-{10\over 9}\beta^2 r^2 }
&{lin.\ (transfer)} 
\cr 
- {  5^{5/2} \over 2^{1/2} 7^{5/2} \pi^{1/2} }\;
{b \beta \over m^2 } \;
\langle \vec {\rm L} \cdot \vec {\rm S} \rangle 
\; e^{-{15\over 14}\beta^2 r^2 }
&{lin.\ (capture)} \ .
}
\end{equation}
The OGE, linear 
and total spin-orbit potentials for the 
$^3$P$_2$ wave
of I=2 $\pi\rho$ 
are shown in Fig.10 for
our standard parameter set.
The largest
contribution to 
the $\pi\rho$ spin-orbit force comes from OGE transfer 
diagrams; the linear confinement spin-orbit from the transfer diagrams
is about half as large and opposite in sign, and the two capture diagram
contributions are much smaller. 
Since the confinement capture and transfer diagrams have opposite
signs, the net result is dominance of the PsV spin-orbit by the 
OGE contribution.

\subsection{Scattering in J$^{{\bf PC}_n}$-Exotic PsV Channels.}

\begin{table}[h]
\caption{
J$^{PC_n}$ Exotic States in PsV.
}
\vskip 0.5cm
\begin{tabular}{cccc|cccccc}
&
\m{3}{c}{Channel}
&
\m{6}{c}{Exotic Quantum Numbers}
\\
\tableline
&
Meson Pair
& 
I$_{tot}$
&
&  S
&  P
&  D
&  F
&  G
&
 \\
\tableline
&
$\pi\rho$
&
0, 2
&
&
$-$
&
$0^{--}$
&
$2^{+-}$
&
$-$
&
$4^{+-}$
&
\\
&
&
1
&
&
$-$
&
$1^{-+}$
&
$-$
&
$3^{-+}$
&
\\
\tableline
&
$\pi\omega$,
$\eta\rho$ 
&
1
&
&
$-$
&
$0^{--}$
&
$2^{+-}$
&
$-$
&
$4^{+-}$
&
\\
\tableline
&
$\eta\omega$
&
0
&
&
$-$
&
$0^{--}$
&
$2^{+-}$
&
$-$
&
$4^{+-}$
&
\\
\end{tabular}
\label{table1}
\end{table}
The recent evidence for 
J$^{PC_n}$-exotic resonances
$\pi_1(1400)$ and $\pi_1(1600)$ \cite{exotics} 
has made the study of scattering
amplitudes in exotic channels especially interesting. 
The 
surprisingly low mass
of 
the $\pi_1(1400)$ in particular
has led to suggestions that it might not be a ``hybrid'' gluonic
excitation, since these are expected at 
$\approx 1.8-2.0$~GeV \cite{hybrid_mass}. 
Another possibility that the $\pi_1(1400)$ 
is a ``multiquark'', 
perhaps a meson-meson bound state 
in a very attractive 
channel. We can test the plausibility of this type of assignment
by calculating meson-meson scattering
amplitudes in the various exotic channels.

The 
exotic channels accessible to the lightest nonstrange PsV meson pairs
are listed in Table I. 
(We do not tabulate light PsPs exotic amplitudes because they are zero
in this model. The PsPs exotic channels are
odd-$l$
$\pi\eta$,
$\pi\eta'$ and
$\eta\eta'$, whereas 
the quark
interchange model 
PsPs scattering amplitudes are even-$l$,
assuming identical spatial wavefunctions.)
We generally expect the largest scattering amplitudes to be in the
lower partial waves. In PsV the P-wave has the first exotics, which are
J$^{PC_n}=0^{--}$ (all channels except I=1  $\pi\rho$) and
J$^{PC_n}=1^{-+}$ (I=1 $\pi\rho$ only).
Calculation of these scattering
amplitudes 
simply requires 
changing the external $q\bar q$ flavor states attached to the
Feynman diagrams of Fig.1.
The results relative to the I=2 $\pi\rho$ case treated in the paper
are summarized in
Table II. 

\begin{table}[h]
\caption{
Overall flavor factors in diagonal PsV scattering.
}
\vskip 0.5cm
\begin{tabular}{cccc|cc}
&
\m{2}{c}{Channel}
&
\m{3}{c}{Relative Amplitude}
\\
\tableline
&
$\pi\rho$
&
I=2
&
&
\hskip1cm
$+1$
&
\\
&
&
1
&
&
\hskip1cm
$0$
\\
&
&
0
&
&
\hskip1cm
$-1/2$
\\
\tableline
&
$\pi\omega$
&
1
&
&
\hskip1cm
$+1/2$
\\
\tableline
&
$\eta\rho$
&
1
&
&
\hskip1cm
$+1/4$
\\
\tableline
&
$\eta\omega$
&
0
&
&
\hskip1cm
$+1/4$
\\
\end{tabular}
\label{table2}
\end{table}

Inspection of the tables shows that the largest exotic
scattering amplitude should be in the I=2 $\pi\rho$
$0^{--}$ P-wave. The elastic phase shift in this channel
is the $^3$P$_0$ curve in Fig.8.
The large negative phase shift shows that this 
is a strongly repulsive channel; the maximum phase shift is predicted
to be a quite large $\approx -50^o$ at $M_{\pi\rho}\approx 3.1$ GeV,
which exceeds even the S-wave phase shift maximum.
The largest {\it attractive} exotic phase shift we have found in PsV is 
the 
I=0 partner, which is $-1/2$ of I=2, giving 
a maximum phase shift of $\approx +25^o$ at the same mass.
We do not find sufficient attraction to form a 
meson-meson ``molecular'' bound state
in any of these nonstrange J$^{PC_n}$-exotic PsV channels. 
The $\eta$ channels are relatively weak because only the
$n\bar n$ part of the $\eta$ contributes to these 
diagonal scattering 
amplitudes; the $s\bar s$ component leads to 
open-strange 
final states
(K$^*{\bar {\rm K}}$ for example) after quark line interchange.

Regarding candidate exotic resonances, there have been 
speculations that
the determination of the mass and width
of the exotic candidate 
$\pi_1(1400)$
may have been compromised
by inelastic rescattering effects \cite{DP}, analogous to the
``Deck effect'' proposed as a nonresonant explanation of the 
$a_1(1260)$. 
For example,
crossing the $\pi b_1$ threshold 
at $\approx 1.4$~GeV
in the process
$\pi\rho \to\pi b_1\to \pi\eta $ might mimic resonant
phase motion if this process has a rapidly varying inelasticity.
We can test this and other nonresonant possibilities 
by calculating the elementary $2\to 2$ 
scattering amplitudes 
using our quark model approach.

Some important results follow from simple flavor factors.
Note in particular that the nonresonant scattering amplitude
$\pi\rho \to \pi\rho$ vanishes in any I=1 channel, including 
the $\pi_1$ exotic one.
This is a general result whenever the 
quark line diagram
of Fig.1 dominates; clearly a pair of oppositely charged, nonstrange 
$q\bar q$ mesons 
$A^+B^-$ 
cannot scatter into another
charged pair $C^+D^-$ under quark interchange. 
A comparison with isospin matrix elements shows that
this implies that scattering
of any two $q\bar q$ isovectors in I=1 vanishes. 
This isospin selection rule 
eliminates two subprocesses
discussed by Donnachie and
Page \cite{DP} as Deck effect backgrounds that might 
shift a higher-mass exotic resonance to an apparent $\pi_1(1400)$,
$\pi\rho \to \pi b_1\to \pi\eta$
and
$\pi\rho \to \pi\rho \to \pi \eta$. 

Independent of any scattering model, one should note that the
coupling $\pi\rho\to\pi b_1$ is probably small because of the 
strong VES experimental limit 
(reported by V.Dorofeev \cite{exotics}) of
\begin{equation}
B(\pi_2(1670)\to \pi b_1) < 0.19\% \hskip1cm   (2\sigma \ c.l.)\ .
\label{pi2}
\end{equation}     
Since $\pi_2 \to \pi\rho$ is a large mode
($B = 31(4) \% $ \cite{PDG98}), if 
$\pi\rho\to\pi b_1$ rescattering were important we would also
expect to observe a large $\pi_2 \to\pi b_1$ branching fraction. 

We also expect the final 
background process
suggested by Donnachie and Page 
($\pi\rho \to \pi\eta \to \pi \eta$) 
to be small,
because the direct time ordering 
$\pi\eta \to \pi \eta$ 
vanishes in P-wave in this model. 
Finally, 
the rescattering process 
they propose,
$\pi_1(1600) \to \pi b_1
\to \eta \pi$, 
does not vanish in the quark interchange 
model,
although the required $\Delta l=1$  and 
suppressed $\eta$ flavor factor may
nonetheless make this a relatively weak amplitude. A calculation of this 
and related
scattering amplitudes is planned for a future publication.

\subsection{Experimental prospects for measuring PsV phase shifts.}

Although there is little experimental information about
PsV
interactions at present, these phase shifts actually are
experimentally accessible in existing data, 
for example as relative FSI phases in the D and S
amplitudes in $b_1\to \omega\pi$. 
These are usually, and incorrectly, taken
to be relatively real amplitudes. The relative phase including the FSI 
is 
${\rm D/S} = |{\rm D/S}| \cdot e^{ i(\delta_D - \delta_S )}$ 
\cite{Watson},  
and is observable 
for example as a
reduction in the strength of
the SD cross term in the $\pi\omega$ angular distribution
by $\cos(\delta_S - \delta_D )$. 
Since this method requires individual measurements of the S$^2$, 
D$^2$ and SD cross term in the angular distribution, it should be 
applicable to cases such as 
$b_1(1230)\to \pi\omega$ and
$b_1(1600)\to \pi\omega$ where S and D are of comparable magnitude
\cite{Pick}. The 
$\delta_{SD} = \delta_S - \delta_D$ 
phases we predict at these
masses 
(which are calculated from +1/2 times the $^3$S$_1$ and  $^3$D$_1$
I=2 $\pi\rho$ phases in Figs.7,8)
are
$\delta_{SD}^{\pi\omega}(M_{\pi\omega}=1.23\; {\rm GeV}) = -14.^o$
and
$\delta_{SD}^{\pi\omega}(M_{\pi\omega}=1.60\; {\rm GeV}) = -17.^o$.

This proposed technique is similar to that used in
K${}_{e4}$ decays \cite{MP}, in which the
low energy I=0 $\pi\pi$ S-wave 
phase shift is actually observed as the difference
between the I=0 S-wave and I=1 P-wave $\pi\pi$ FSI phases.

\section{Summary and Conclusions}

In this paper we have derived meson-meson scattering amplitudes, including
spin-dependent forces, from a 
calculation of the 
Born-order 
matrix element of the quark-quark
interaction between two-meson states. 
Since $q\bar q$ annihilation
is not included in the model, it 
describes scattering that does not involve
coupling to $s$-channel resonances. 
This includes for example 
I=2 and the nonresonant backgrounds in all channels,
including exotic J$^{PC_n}$.

We considered the cases of PsPs and PsV scattering, and derived
the scattering amplitudes in all $j,l,s$ channels for these cases.
The parameters of the model were
previously fixed by quark model studies of hadron spectroscopy. 
Where 
possible we have compared the results to experiment.

In I=2 $\pi\pi$ (the best studied PsPs case) the results were
shown to be in reasonable agreement with experiment in S-wave scattering,
and an extremum predicted near $M_{\pi\pi}=1.5$~GeV is supported
by the data. 
Rapid variation of $\delta_0/k_\pi$ is predicted near threshold,
which may reconcile theoretical expectations of a 
small scattering length 
with larger reported 
experimental values based on extrapolation in ${k_\pi}^2$.
The experimental D-wave, although quite small,
is clearly larger than the model 
predicts.

The PsV system is a convenient theoretical laboratory for studying
spin-dependent forces, since it can accommodate both spin-orbit and
tensor interactions, and is simpler than KN or NN. We
derived analytical results for these spin-dependent
PsV interactions (T-matrices and phase shifts) given SHO wavefunctions and the
standard spin-dependent quark model forces. 
The quark-quark spin-orbit and tensor forces map directly into spin-orbit
and tensor PsV interactions.
We find that the OGE
spin-orbit force in the PsV system is quite large in P-wave, and so
is expected to be large in many other hadron-hadron systems as well. 

There is no PsV phase shift data at present. 
We noted however that PsV phase shifts
actually can be measured in multiamplitude resonance decays to PsV
final states, so it should be possible to test theoretical
predictions for PsV scattering amplitudes in future experimental 
studies. 

Our predictions for scattering in J$^{PC_n}$-exotic channels are
of current interest  
because 
the reported exotics might 
be complicated by large and rapidly varying 
nonresonant inelasticities.
One speculation is that the 
$\pi_1(1400)$ parameters might be strongly affected by
the opening of inelastic couplings to the $\pi b_1$ channel.
In our model (and in any 
$q\bar q$ constituent
interchange model) several of these nonresonant processes 
can be
rejected as significant complications
because of
vanishing flavor factors.

In future we plan to extend our calculations to other exotic
meson-meson channels, such as S+P, to test whether strong 
attractive interactions 
are predicted
that might support ``multiquark exotics" such as S+P molecules.
We also plan to 
apply the current 
approach to the study of spin-dependent interactions in 
other hadronic systems, including KN, NN and
light hadron + charmonium systems.

\section{Acknowledgements}

We are grateful to our colleagues
D.Blaschke,
N.Cason,
H.G.Dosch,
S.Krewald,
E.Klempt,
N.Isgur,
C.Michael,
M.R.Pennington,
B.Pick,
H.-J.Pirner,
G.R\"opke,
F.Sassen, 
J.Speth, 
Fl.Stancu 
and
C.-Y.Wong
for useful discussions of various issues relating to meson-meson scattering.

This research was supported in part by the DOE Division of Nuclear Physics,
at ORNL,
managed by UT-Battelle, LLC, for the US Department of Energy 
under Contract No. DE-AC05-00OR22725.
ES acknowledges support from the DOE under grant
DE-FG02-96ER40944 and DOE contract DE-AC05-84ER40150 under which the
Southeastern Universities Research Association operates the Thomas
Jefferson National Accelerator Facility. TB acknowledges additional
support from the Deutsche Forschungsgemeinschaft DFG 
at the University of Bonn and the Forschungszentrum J\"ulich 
under contract Bo
56/153-1. 

\newpage

\renewcommand{\theequation}{A\arabic{equation}}
\setcounter{equation}{0}  
\section*{Appendix A: Quark-level T-matrices and Wavefunctions}  

The various contributions to the 
quark-quark 
$T_{fi}$
(with color factors of $T^aT^a$ removed)
are 
\\
\begin{equation}
T_{fi}(\vec q, \vec p_1, \vec p_2\, ) 
=
\cases{
-{8\pi \alpha_s \over  3 m^2 }\, \bigg[ \vec S_1\cdot \vec S_2\bigg] 
&{\rm OGE spin-spin}\cr 
+{4\pi \alpha_s \over   \vec q^{\, 2}}
&{\rm OGE color Coulomb}\cr 
+{6\pi b \over \vec q^{\, 4} }
&{\rm linear conft.} \cr 
+{4i\pi \alpha_s \over  m^2\vec q^{\, 2}}\;
\bigg[
\vec S_1\cdot \Big( \vec q\times ({\vec p_1\over 2}-\vec p_2)\Big) +
\vec S_2\cdot \Big(\vec q\times (\vec p_1-{\vec p_2\over 2} ) \Big) 
\bigg]
&{\rm OGE spin-orbit}\cr 
-
{3i\pi  b \over m^2\vec q^{\, 4}}\;
\bigg[
\vec S_1\cdot ( \vec q\times \vec p_1 ) -
\vec S_2\cdot ( \vec q\times \vec p_2 )
\bigg]
&{\rm linear spin-orbit} \cr
+{4\pi \alpha_s \over m^2\vec q^{\; 2}}\;
\bigg[
\vec S_1 \cdot \vec q \;
\vec S_2 \cdot \vec q 
-{1\over 3}\;
\vec q^{\; 2}\,
\vec S_1\cdot \vec S_2
\bigg]
&{\rm OGE tensor.}\cr 
} 
\label{quarkT}
\end{equation}  

The standard $q\bar q$ 
quark model Gaussian wavefunction is given by
\begin{equation}
\Phi(\vec p_{rel}) =
{1 \over \pi^{3/4} \beta^{3/2} } \;
e^{-\vec p_{rel}^{\; 2} / 8 \beta^2 } 
\end{equation}  
where in general
\begin{equation}
\vec p_{rel} =
{ 
m_{\bar q} \vec p_q - 
m_q \vec p_{\bar q} 
\over 
(m_q + 
m_{\bar q} )/2 
} 
\end{equation}  
and for our special case of equal quark and antiquark masses
\begin{equation}
\vec p_{rel} =
\vec p_q - 
\vec p_{\bar q} \ .  
\end{equation}  

\newpage
\renewcommand{\theequation}{B\arabic{equation}}
\setcounter{equation}{0}  
\setcounter{subsection}{0}  
\section*{Appendix B: Explicit Overlap Integrals}

\subsubsection*{B1. Results Included}

In this appendix we give the explicit
meson-meson T-matrix elements
that follow from the overlap integrals 
Eqs.(\ref{T1def}-\ref{C2def}) 
with
Gaussian wavefunctions and the various quark T-matrix elements.
The 
OGE spin-spin hyperfine, 
color Coulomb and
linear 
confinement 
results were derived previously 
\cite{BB}. For completeness we quote the formulas here, as well
as giving the new spin-orbit and tensor results.
The multiplicative diagram-dependent color and flavor 
factors and the signature
phase (which is $(-1)$ for these meson-meson scattering diagrams) are
not included in the results given below. 
These formulas abbreviate
the confluent hypergeometric function
${}_1{\rm F}_1(a;c;x)$
as
${\rm f}_{a,c}(x)$,
and 
$\vec Q_\pm = (\vec C \pm \vec A \, )$.

\subsubsection*{B2. OGE Spin-Spin Hyperfine Contribution}
These simple contact matrix elements were evaluated previously, 
for example in
Ref.~\cite{pipi} (in an equivalent form, but incorporating color factors
and the signature phase, as Eqs.(71-73) of that reference). 
The results are
\begin{eqnarray}
& &
{T_{fi}}^{\rm (T1)} = 
-\;
{2^3 \over 3}\;
{\pi\alpha_s \over m^2}\;
e^{-\vec Q_+^2/8\beta^2}
\bigg[
\vec S_1\cdot \vec S_2
\bigg]
\\
& &
{T_{fi}}^{\rm (T2)} = 
{T_{fi}}^{\rm (T1)}(\vec C \rightarrow -\vec C) 
\\
\rule{0mm}{1cm}
& &
{T_{fi}}^{\rm (C1)} = 
-\;
{2^6 \over 3^{5/2}}\;
{\pi\alpha_s \over m^2}\;
e^{-\vec A^{\, 2}/3\beta^2}
\bigg[
\vec S_1\cdot \vec S_2
\bigg]
\\
& &
{T_{fi}}^{\rm (C2)} = 
{T_{fi}}^{\rm (C1)} 
\ . 
\end{eqnarray}

\subsubsection*{B3. OGE Color Coulomb Contribution}
The contribution of the OGE color Coulomb interaction to the meson-meson 
T-matrix follows from the evaluation of the integrals 
Eqs.(\ref{T1def}-\ref{C2def}) 
with the
second quark-quark $T_{fi}$ in Eq.(\ref{quarkT}). The results are

\begin{eqnarray}
\rule{0mm}{1cm}
& &
{T_{fi}}^{\rm (T1)} = 
+\;
{2^2 }\;
{\pi\alpha_s \over \beta^2}\;
{\rm f}_{{1\over 2},{3\over 2}}(\vec Q_-^{\, 2} /8\beta^2) 
\;
e^{-\vec A^{\, 2}/2\beta^2}
\label{B5}
\\
& &
{T_{fi}}^{\rm (T2)} = 
{T_{fi}}^{\rm (T1)}(\vec C \rightarrow -\vec C) 
\\
\rule{0mm}{1cm}
& &
{T_{fi}}^{\rm (C1)} = 
+\;
{2^3 \over 3^{1/2}}\;
{\pi\alpha_s \over \beta^2}\;
{\rm f}_{{1\over 2},{3\over 2}}(\vec A^{\, 2} /6\beta^2) 
\;
e^{-\vec A^{\; 2}/2\beta^2}
\label{B7}
\\
& &
{T_{fi}}^{\rm (C2)} = 
{T_{fi}}^{\rm (C1)}
\ . 
\end{eqnarray}

\subsubsection*{B4. Linear Confinement Contribution}
The linear confinement 
integrals were carried out in coordinate space,
since the Fourier transform of the linear potential is 
singular. 
The results are
\begin{eqnarray}
& &
{T_{fi}}^{\rm (T1)} = 
-6\,
{\pi b\over \beta^4}\;
{\rm f}_{-{1\over 2},{3\over 2}}(\vec Q_-^{\, 2} /8\beta^2) 
\;
e^{-\vec A^{\, 2}/2\beta^2}
\\
& &
{T_{fi}}^{\rm (T2)} = 
{T_{fi}}^{\rm (T1)}(\vec C \rightarrow -\vec C) 
\\
\rule{0mm}{1cm}
& &
{T_{fi}}^{\rm (C1)} = 
-\;
{3^{3/2} }\;
{\pi b\over \beta^4}\;
{\rm f}_{-{1\over 2},{3\over 2}}(\vec A^{\, 2} /6\beta^2) 
\;
e^{-\vec A^{\, 2}/2\beta^2}
\\
& &
{T_{fi}}^{\rm (C2)} = 
{T_{fi}}^{\rm (C1)} 
\ . 
\label{B12}
\end{eqnarray}
One may also obtain these results using the momentum space
integrals 
Eqs.(\ref{T1def}-\ref{C2def}), 
but 
the $1/q^4$ quark-quark $T_{fi}$
in Eq.(\ref{quarkT}) must include
a long-distance regularization 
in the intermediate stages of the integration.
The final result is well defined due to the Gaussian damping provided
by the
wavefunctions. 
We have checked both the linear and color
Coulomb 
$T_{fi}$ results by comparing the expressions Eqs.(\ref{B5}-\ref{B12})
with Monte Carlo evaluations
of the corresponding real-space overlap integrals.

\subsubsection*{B5. OGE Spin-Orbit Contribution}
The four OGE 
spin-orbit overlap integrals can be evaluated similarly using the 
fourth quark-quark $T_{fi}$ in Eq.(\ref{quarkT}), which gives 

\begin{eqnarray}
& &
{T_{fi}}^{\rm (T1)} = 
-
{\pi\alpha_s\over m^2\beta^2}\;
{\rm f}_{{3\over 2},{5\over 2}}(\vec Q_-^{\, 2} /8\beta^2) \;
e^{-\vec A^{\, 2}/2\beta^2}
\;
\bigg[
(\vec S_1 + \vec S_2)\cdot i(\vec A \times \vec C)
\bigg]
\\
& &
{T_{fi}}^{\rm (T2)} = 
{T_{fi}}^{\rm (T1)}(\vec C \rightarrow -\vec C) 
\\
\rule{0mm}{1cm}
& &
{T_{fi}}^{\rm (C1)} = 
+{4 \over 3^{5/2} }\;
{\pi\alpha_s\over m^2\beta^2}\;
{\rm f}_{{3\over 2},{5\over 2}}(\vec A^{\, 2} /6\beta^2) 
\;
e^{-\vec A^{\, 2}/2\beta^2}\;
\bigg[
(\vec S_1 - \vec S_2)\cdot i(\vec A \times \vec C )
\bigg]
\\
& &
{T_{fi}}^{\rm (C2)} = 
-\; {T_{fi}}^{\rm (C1)} 
\ . 
\end{eqnarray}

\subsubsection*{B6. Scalar Confinement Spin-Orbit Contribution}
The matrix elements 
Eqs.(\ref{T1def}-\ref{C2def}), 
of the scalar confinement
spin-orbit interaction in Eq.(\ref{quarkT}) are

\begin{eqnarray}
& &
{T_{fi}}^{\rm (T1)} = 
+{1  \over 2}\;
{\pi b\over m^2\beta^4}\;
{\rm f}_{{1\over 2},{5\over 2}}(\vec Q_-^{\, 2} /8\beta^2) \;
e^{-\vec A^{\, 2}/2\beta^2}\;
\bigg[
(\vec S_1 + \vec S_2)\cdot i (\vec A \times \vec C)
\bigg]
\\
& &
{T_{fi}}^{\rm (T2)} = 
{T_{fi}}^{\rm (T1)}(\vec C \rightarrow -\vec C) 
\\
\rule{0mm}{1cm}
& &
{T_{fi}}^{\rm (C1)} = 
+{1  \over 2\cdot 3^{1/2} }\;
{ \pi b\over m^2\beta^4}\;
{\rm f}_{{1\over 2},{5\over 2}}(\vec A^{\, 2} /6\beta^2) 
\;
e^{-\vec A^{\, 2}/2\beta^2}\;
\bigg[
(\vec S_1 - \vec S_2)\cdot i (\vec A \times \vec C)
\bigg]
\\
& &
{T_{fi}}^{\rm (C2)} = 
-\; {T_{fi}}^{\rm (C1)} 
\ . 
\end{eqnarray}

\subsubsection*{B7. OGE Tensor Contribution}
Finally, for the OGE tensor $T_{fi}$ 
(the last entry in Eq.(\ref{quarkT}) we find 

\begin{eqnarray}
& &
{T_{fi}}^{\rm (T1)} = 
+\;
{1\over 5 }\;
{\pi\alpha_s\over m^2\beta^2}\;
{\rm f}_{{5\over 2},{7\over 2}}(\vec Q_-^{\, 2} /8\beta^2)\;  
e^{-\vec A^2/2\beta^2}\;
\bigg[
\vec S_1 \cdot \vec Q_- \;
\vec S_2 \cdot \vec Q_- 
-{1\over 3}\,
\vec Q_-^{\, 2}\;
\vec S_1 \cdot \vec S_2
\bigg]
\\
& &
{T_{fi}}^{\rm (T2)} = 
{T_{fi}}^{\rm (T1)}(\vec C \rightarrow -\vec C) 
\\
& &
{T_{fi}}^{\rm (C1)} = 
+\;
{2^5 \over 3^{5/2}\cdot 5}\;
{\pi\alpha_s\over m^2\beta^2}\;
{\rm f}_{{5\over 2},{7\over 2}}(\vec A^{\, 2} /6\beta^2)\; 
e^{-\vec A^{\, 2}/2\beta^2}\;
\bigg[
\vec S_1 \cdot \vec A\;
\vec S_2 \cdot \vec A
-{1\over 3}\,
\vec A^{\, 2}\;
\vec S_1 \cdot \vec S_2
\bigg]
\label{B23}
\\
& &
{T_{fi}}^{\rm (C2)} = 
{T_{fi}}^{\rm (C1)} 
\label{B24}
\ . 
\end{eqnarray}

The tensor matrix elements in the
capture diagrams, Eqs.(\ref{B23},\ref{B24}), are 
the only cases in which we have found a
post-prior discrepancy in these PsPs and 
PsV scattering amplitudes; 
the post forms of these matrix elements 
involve a tensor in $\vec C$ rather than $\vec A$, 
\begin{equation}
{T_{fi}}^{\rm (C1, post)} = 
+\;
{2^5 \over 3^{5/2}\cdot 5}\;
{\pi\alpha_s\over m^2\beta^2}\;
{\rm f}_{{5\over 2},{7\over 2}}(\vec A^{\, 2} /6\beta^2) \;
e^{-\vec A^{\, 2}/2\beta^2}\;
\bigg[
\vec S_1 \cdot \vec C\;
\vec S_2 \cdot \vec C
-{1\over 3}\,
\vec C^{\, 2}\;
\vec S_1 \cdot \vec S_2
\bigg] \ .
\end{equation}
These capture tensor terms vanish in the PsPs channel. They do 
make a small, off-diagonal 
contribution to
PsV scattering, {\it albeit} only 
in the ${}^3$S$_1 \leftrightarrow  {}^3$D$_1$ amplitude.

\newpage
\renewcommand{\theequation}{C\arabic{equation}}
\setcounter{equation}{0}  
\setcounter{subsection}{0}  
\section*{Appendix C: Mapping quark spins into hadron spins.}

\subsubsection*{C1. Spin Matrix Elements}

In these scattering amplitude calculations the matrix elements of spin-dependent
quark interactions (the spin-spin, spin-orbit and tensor forces) 
involve matrix elements
of linear and bilinear 
quark spin operators. Since the quark spins are not directly observed, it is
useful to replace them by the spins
of the external hadrons. This appendix gives the (diagram dependent)
mapping
from quark spins to hadron spins 
in the PsPs and PsV cases considered in this paper.

The spin matrix elements we require are $I$, 
$S(1)^i$, 
$S(2)^i$ 
and
$S(1)^i 
S(2)^j$ 
between general initial and final PsPs and PsV spin states.
Our 
convention for the diagrams (Fig.1) is that mesons 
A and C are always Ps 
({\it e.g.} $\pi$), 
and
B and D are Ps or V
({\it e.g.} $\pi$ or $\rho$). 

\subsubsection*{C2. PsPs}

First, in PsPs scattering there are no external meson spins, so the 
quark spin matrix elements
are proportional to geometrical tensors such as $\delta^{ij}$. The matrix
elements
by diagram 
are
\begin{eqnarray}
& &
\langle 
PsPs|
\,
I
\,
|
PsPs
\rangle 
= +{1\over 2} 
\qquad 
\mbox{all diagrams}
\\
& &
\nonumber\\
& &
\langle 
PsPs|
\,
S(1)^i
\,
|
PsPs
\rangle 
= 
\langle PsPs|
\,
S(2)^i
\,
|
PsPs
\rangle 
= 
0 
\qquad 
\mbox{all diagrams}
\\
& &
\nonumber\\
& &
\langle 
PsPs|
\,
S(1)^i  
S(2)^j
\,
|
PsPs
\rangle 
=
\cases{  
+{1\over 8}\; \delta^{ij} 
&
{\rm T1, T2, T1$_{symm}$, T2$_{symm}$} \cr
-{1\over 8}\; \delta^{ij} 
&
{\rm C1, C2, C1$_{symm}$, C2$_{symm}$}\cr
}
\qquad 
\mbox{hence}
\\
& &
\nonumber\\
& &
\langle 
PsPs|
\,
\vec S(1) \cdot 
\vec S(2)
\,
|
PsPs
\rangle 
=
\cases{  
+{3\over 8}  
& 
{\rm T1, T2, T1$_{symm}$, T2$_{symm}$} 
\cr
-{3\over 8}  
&
{\rm C1, C2, C1$_{symm}$, C2$_{symm}$} 
\cr
} \ .
\end{eqnarray}
Note that 
the spin-orbit and tensor terms 
vanish identically in
PsPs scattering; this follows
from 
applying 
Eqs.(C2,C3) 
to Eq.(A1).  

\subsubsection*{C3. PsV}

In PsV scattering the 
vector ({\it e.g.} $\rho$) meson spin $\vec S_\rho $ provides an
additional degree of freedom, and the 
linear and quadratic quark spin matrix elements can be expressed
in terms of 
the
$\rho$ spin matrix elements
$\langle \rho_f | S_\rho^i | \rho_i \rangle $
and
$\langle \rho_f | S_\rho^i S_\rho^j  | \rho_i \rangle $.
The mapping of quark to meson spins is

\begin{eqnarray}
& &
\langle (\pi\rho)_f | 
I 
| (\pi\rho)_i \rangle  
=
+{1 \over 2}\, \langle \rho_f |I|\rho_i \rangle
\ \ \ \ \
{\rm all \ diagrams } 
\\
& & \nonumber \\
& &
\langle (\pi\rho)_f | 
S(1)^i 
| (\pi\rho)_i \rangle  
=
\cases{
-{1 \over 4}\, \langle \rho_f |S_\rho^i|\rho_i \rangle
&{\rm T1, T2$_{symm}$, C1, C2$_{symm}$ }\cr 
+{1 \over 4}\, \langle \rho_f |S_\rho^i|\rho_i \rangle
&{\rm T2, T1$_{symm}$, C2, C1$_{symm}$ }\cr 
} 
\\
& & \nonumber \\
& &
\langle (\pi\rho)_f | 
S(2)^i 
| (\pi\rho)_i \rangle  
=
+{1 \over 4}\, \langle \rho_f |S_\rho^i|\rho_i \rangle
\ \ \ \ \
{\rm all \ diagrams } 
\\
& & \nonumber \\
& &
\langle (\pi\rho)_f | 
S(1)^i 
S(2)^j 
| (\pi\rho)_i \rangle  
=
\nonumber
\\
& & \nonumber \\
& &
\cases{
+{1\over 8}\, \delta^{ij}
\langle \rho_f |I|\rho_i \rangle
+{1 \over 8}\, 
i\, \epsilon^{ijk}\, \langle \rho_f |S_\rho^k|\rho_i \rangle
-{1 \over 4}\, \langle \rho_f |S_\rho^i S_\rho^j|\rho_i \rangle
&{\rm T1, T2$_{symm}$ }\cr 
+{1\over 8}\, \delta^{ij}
\langle \rho_f |I|\rho_i \rangle
+{1 \over 8}\, 
i\, \epsilon^{ijk}\, \langle \rho_f |S_\rho^k|\rho_i \rangle
&{\rm T2, T1$_{symm}$ }\cr 
-{1\over 8}\, \delta^{ij}
\langle \rho_f |I|\rho_i \rangle
-{1 \over 8}\, 
i\, \epsilon^{ijk}\, \langle \rho_f |S_\rho^k|\rho_i \rangle
&{\rm C1,  C2$_{symm}$ }\cr 
-{1\over 8}\, \delta^{ij}
\langle \rho_f |I|\rho_i \rangle
-{1 \over 8}\, 
i\, \epsilon^{ijk}\, \langle \rho_f |S_\rho^k|\rho_i \rangle
+{1 \over 4}\, 
\langle \rho_f |S_\rho^i S_\rho^j|\rho_i \rangle
&{\rm C2,  C1$_{symm}$ }\cr 
} \ . 
\end{eqnarray}

\newpage
\renewcommand{\theequation}{D\arabic{equation}}
\setcounter{equation}{0}  
\section*{Appendix D: Phase Shifts and Inelasticities from the T-matrix}

Since total angular momentum is conserved, the 
T-matrix is block
diagonal in a total angular momentum basis,
and can be written as
\begin{equation}
{\rm T} = 
\sum_{jm} 
\;
| 
jm
\rangle 
\;
T_{j}
\;
\langle 
jm
|  \ .
\end{equation}
The coefficients $\{ T_j \}$ can be determined by evaluating 
the matrix element
\begin{equation}
T_j = 
\;
\langle  
jm
| 
\;
{\rm T}
\;
| 
jm
\rangle \ . 
\end{equation}
In the special case of spinless scattering these basis states are
eigenstates of orbital angular momentum
\begin{equation}
| l m \rangle = 
\int d\Omega \; 
Y_{l m}(\Omega) \;
| 
\Omega 
\rangle  \ ,
\end{equation}
so the T-matrix is given by

\begin{equation}
{\rm T} = 
\sum_{l m} 
\;
| 
l m 
\rangle 
\,
T_{l}   
\,
\langle 
l m 
| 
=
\sum_{l m} 
\int \!\!\! 
\int  
d\Omega \; 
d\Omega' \  
| 
\Omega' 
\rangle \
Y_{l m}(\Omega') \;
T_l \;
Y^*_{l m}(\Omega) \;
\langle 
\Omega 
|   \ .
\end{equation}
The T-matrix 
can also be written in terms of the scattering amplitude 
$ T_{fi}(\Omega, \Omega') $ 
between momentum
eigenstates,
\begin{equation}
{\rm T} =
\int \!\!\! 
\int  
d\Omega \; 
d\Omega' \  
| 
\Omega' 
\rangle \
T_{fi}(\Omega, \Omega') \  
\langle 
\Omega 
| \ , 
\end{equation}
so $ T_{fi}(\Omega, \Omega') $ and $T_l$ are related by
\begin{equation}
T_{fi}(\Omega, \Omega') \  
=
\sum_{l} 
T_l \;
\sum_{m} 
Y_{l m}(\Omega') \;
Y^*_{l m}(\Omega) 
=
\sum_{l} 
{2l + 1\over 4\pi }\; T_l \;
P_l(\mu)  
\label{TP}
\end{equation}
and hence
\begin{equation}
T_l = 
\;
\langle  
l m 
| 
\,
{\rm T}
\,
| 
l m 
\rangle 
=
\int \!\!\! 
\int  
d\Omega \, 
d\Omega' \  
Y^*_{l m}(\Omega') \;
T_{fi}(\Omega, \Omega')\;   
Y_{l m}(\Omega) \;
\end{equation}
where $\mu = 
\cos \theta_{\Omega \Omega' }$. 
A more familiar quantum-mechanical result follows from 
fixing
the incident direction $\Omega=\hat z$ 
in Eq.(\ref{TP})
and integrating over final
angles with a $P_l(\mu)$ weight, which gives 
\begin{equation}
T_l = 2\pi \; \int_{-1}^1 \! d\mu \; P_l(\mu)\; 
T_{fi}(\hat z,\Omega') \ .
\label{Tl}
\end{equation}

Since we define 
$T_{fi}(\Omega, \Omega')$ by
\begin{equation}
\langle C_{\Omega'}, D_{-\Omega'}\, |
\, S
\,
|A_\Omega, B_{-\Omega}\, \rangle 
=
\delta_{fi} - i\; (2\pi)^4\; \delta^{(4)}(A + B - C - D)\; T_{fi} \ ,
\end{equation}
it is related to the Lorentz invariant $2\to 2$ scattering amplitude 
${\cal M}$ defined
by the PDG\cite{PDG98} (their Eq.(35.8)) by
\begin{equation}
T_{fi} = 
{
{\cal M} 
\over 
\prod_{n=1}^4 (2E_n)^{1/2} 
}
\end{equation}
and hence to the CM scattering amplitude $f(k,\theta)$ (their Eq.(35.48)) by
\begin{equation}
T_{fi} = -{8\pi \over \sqrt{s}}\; f(k,\theta) \ .
\label{Tf}
\end{equation}
(Here $k$ is the CM momentum of any particle,
$k=
|\vec A | =
|\vec B | =
|\vec C | =
|\vec D |$.)
The partial wave expansion of $f(k,\theta)$, Eq.(35.44),
\begin{equation}
f(k,\theta) =
\sum_l \; (2l+1)\, a_l \, P_l(\mu)
\end{equation}
and the relation between a diagonal 
partial wave amplitude $a_l$ and the phase shift
\begin{equation}
a_l = {e^{2i\delta_l} -1 \over 2i}
\label{al}
\end{equation}
allow us to determine $\delta_l$ from $T_{fi}$. For purely
elastic scattering,
and assuming small 
phase shifts so that
$a_l\approx \delta_l$, Eqs.(\ref{TP}-\ref{al}) give 
\begin{equation}
\delta_l = 
-{1 \over 4\pi} \; 
{k E_A E_B\over \sqrt{s}}  \; 
\int \! d\mu \; P_l(\mu)\;
T_{fi}(\hat z,\Omega') \;
=
-{1 \over 8\pi^2} \; 
{k E_A E_B\over \sqrt{s}}  \; 
T_l \ .
\label{deltal1}
\end{equation}
This relation was used previously 
to determine 
for example 
K$\pi$ 
\cite{Kpi} and
$l$-diagonal KN
\cite{KN}
elastic phase shifts.
In the case of elastic scattering of identical bosons, 
such as
I=2 $\pi\pi$, 
there is an additional factor of two for identical particles
\cite{Watson}, so
the relation between the Born-order
phase shift and 
the T-matrix element becomes
\begin{equation}
\delta_l\bigg|_{ident.} = -{ k E_A \over 16\pi} \int^{+1}_{-1}\, 
T_{fi}\,
P_{\ell}(\mu) \, d\mu \ .
\end{equation}

Since the angular integral in Eq.(\ref{deltal1}) is proportional to
the amplitude $\langle lm |\, T \, | lm \rangle$, 
we may also write the Born-order elastic phase shift 
(for distinguishable particles) directly in
terms of the T-matrix,
\begin{equation}
\delta_l = 
-{1 \over 8\pi^2}\;
{k E_A E_B\over \sqrt{s} }  \ 
\langle lm |\, T \, | lm \rangle  \ .
\end{equation}

This formula has a straightforward generalization
to the case of external hadrons
with spin, which we use to evaluate $\pi\rho$ phase shifts and inelasticities.
In the case of an $l$- and $s$-diagonal interaction this is
\begin{equation}
\delta_{jls}  = 
-{1 \over 8\pi^2}\;
{k E_A E_B\over \sqrt{s} }  \ 
\langle jm;ls |\, T \, | jm;ls \rangle  \ .
\end{equation}
This is adequate for diagonal 
forces such as our $\pi\rho$ spin-orbit interactions. The tensor force
however is not $l$-diagonal, so for this case we must introduce a more general
parametrization.
Since the tensor interaction couples $l,l'$ channel pairs which have 
the same
$j$ but differ by 
$|l-l'|=2$, it leads to 
$2\times 2$ ${\cal S}$-matrices. These can be parametrized
as

\begin{equation}
{\cal S}^j =
\left[
\begin{array}{cc}
\eta_{ll'}\ e^{2i\delta_l} 
& 
i\sqrt{1-\eta_{ll'}^2}\ e^{i(\delta_l + \delta_{l'})} 
\\
i\sqrt{1-\eta_{ll'}^2}\ e^{i(\delta_l + \delta_{l'})} 
& 
\eta_{ll'}\ e^{2i\delta_{l'}} 
\end{array}
\right]
\ .
\end{equation}

In our 
calculation, both of the phase shifts and 
the inelasticity $\epsilon_{ll'} \equiv \sqrt{1-\eta_{ll'}^2}$
are $O(H_I)$, 
so to this order we can relate these 
linearly to the matrix elements of $H_I$.
The Born order phase shift formula (D17) remains valid
for both channels,
and the inelasticity is
\begin{equation}
\epsilon_{ll'} 
=
-{1 \over 4\pi^2}\;
{k E_A E_B\over \sqrt{s} }  \ 
\langle jm;l's |\, T \, | jm;ls \rangle  \ .
\end{equation}
The overall phase of $\epsilon_{ll'}$
is dependent on the state
normalizations, but the familiar
\begin{equation}
\eta_{ll'} = \Big| \sqrt{1-\epsilon_{ll'}^2}\; \Big|
\end{equation}
is unique.

\newpage
\renewcommand{\theequation}{E\arabic{equation}}
\setcounter{equation}{0}  
\section*{Appendix E: Potentials from the T-matrix}

Potentials provide 
a very useful representation of hadron-hadron interactions.
They have a clear intuitive meaning, and can easily be used 
in the nonrelativistic Schr\"odinger 
equation in searches for bound states or in coupled channel problems.

Unfortunately, 
one may define hadron-hadron potentials in many different ways. Ideally they
should reproduce phase shifts or T-matrix elements, at least in the low
energy limit. The assumption of a unique, purely local potential 
is in general overly restrictive, as it 
leads to a scattering amplitude that
is a function of $t$ only. In general we find 
$2\to 2$ scattering amplitudes
that depend on both $s$ and $t$. 
One approach to this problem is to 
introduce
nonlocal ``gradient" corrections to the potential \cite{BG},
which can be expressed for example as $V(r)\, 
\vec {\rm L} \cdot \vec {\rm S}$ 
terms; 
this approach leads 
to 
the familiar Breit-Fermi Hamiltonian for one-photon and one-gluon exchange,
and was used to define hadron-hadron potentials in our previous work
\cite{pipi,BB}. 

Alternatively one may project the scattering amplitude 
onto a given angular channel $l$ so
that only $s$ dependence remains, and find a local 
$l$-wave potential that describes
the scattering in that channel. This definition of potentials 
was discussed by Mott and Massey
\cite{MM}, and is equivalent to the definition we shall use here.
This approach was previously used by Swanson \cite{S_annphys} to define
meson-meson potentials from scattering amplitudes.

The quantum mechanical relation between the 
phase shift $\delta_l(k)$ and
the radial wavefunction $R_l(r)$ in potential scattering of a mass $\mu$ 
particle (which becomes the reduced mass below) 
in first Born approximation is
\begin{equation}  
\delta_l = -2 \mu k \int_0^\infty \!\!\! r^2 \, dr \; V(r) 
\, j_l(kr)^2 \ .
\end{equation}
Since our T-matrix elements implicitly determine 
the elastic scattering phase shifts, for example
the $I=2$ $\pi\pi$ S-wave in Eq.(6), we can invert this formula 
for each $l$ to determine the
corresponding $l$-wave local potential $V_l(r)$. In practice we find
that our phase shifts are sufficiently ``hard" at high energies to require 
singular potentials; this is presumably an artifact of
our approximations, such as assuming a contact spin-spin interaction. For this
reason we do not completely invert the phase shift relation (E1), 
and instead 
simply fit an assumed Gaussian form 
\begin{equation}
V(r) = V_g \; e^{-r^2/r_g^2} 
\end{equation}
to our theoretical {\it low energy} phase shift.
The two Gaussian parameters are determined from 
the $O(k)$ and $O(k^3)$ terms in the expansion of the phase shift near
threshold,
which in the S-wave case are equated to 
\begin{equation}
\lim_{k\to 0} \; \delta_0(k) 
= -{{\pi}^{1/2} \over 2} \; \mu \, V_g \, r_g^3 \;
k\; \bigg( 1 - {1\over 2} \; r_g^2\, k^2 + O(k^4) \bigg) \ .
\end{equation}
The generalization to higher $l$, also using Eqs.(E1), is
\begin{equation}
\lim_{k\to 0} \ \delta_l (k) 
= -{{\pi}^{1/2} \over 2} \; \mu \, V_g \, r_g^3 \ 
{(r_g^2 / 2)^l  \over (2l + 1)!! }\ 
k^{2l + 1}  
\;
\bigg( 1 - {1\over 2} \; r_g^2\, k^2 
+ 
O(k^4) \bigg) \ .
\end{equation}

In determining the Gaussian potentials that correspond to our derived
phase shifts such as Eqs.(6,14,28), we
set the 
external factors of
$E_\pi$ 
and
$E_\rho$ 
equal to 
$m_\pi$ 
and
$m_\rho$ 
before expanding
in $k$. This corresponds to 
using nonrelativistic phase space and nonrelativistic external hadron 
line normalizations
in our T-matrix calculations, which we 
assume is the appropriate choice for the derivation of a
nonrelativistic
equivalent potential. 

\newpage
\renewcommand{\theequation}{F\arabic{equation}}
\setcounter{equation}{0}  
\section*{Appendix F: The Post-Prior Discrepancy}

The ``post-prior
discrepancy" is a
familiar problem in rearrangement collisions;
the diagrams of Fig.1 treat the 
scattering as due to an interaction between the initial
hadrons A and B
(``prior"),
but we could equally well have written the interaction 
between the two final hadrons C and D
(``post"). 
The 
post T-matrices may be obtained from the 
prior ones by exchanging the initial and final mesons and transforming
the momenta 
$\vec A \to \vec C$,
$\vec C \to \vec A$ and
$\vec q \to -\vec q$.
Thus for example the 
post C1 T-matrix is
\begin{eqnarray}
& &
T_{fi}^{\rm (C1, post)}(AB\to CD)  = 
\nonumber
\\
& &
\hskip 1cm
\int\! \! \! \int  d^3 q \, d^3p \
\Phi_C^*(2\vec p + \vec q -  \vec C\, ) \;
\Phi_D^*(2\vec p + \vec q -2\vec A - \vec C \, )  
\nonumber
\\
& &
\hskip 1cm
T_{fi}(\vec q, \vec p, - \vec p + \vec A \, ) \  
\Phi_A(2\vec p - \vec q -  \vec A\, ) \;
\Phi_B(2\vec p + \vec q -  \vec A -2\vec C\, ) \ .
\end{eqnarray}
\begin{equation}
= T_{fi}^{\rm (C1, prior)}(AB\to CD)
\bigg|_{
{
(\Phi_A,\Phi_B,\Phi_C^*,\Phi_D^*)\to(\Phi_C^*,\Phi_D^*,\Phi_A,\Phi_B),
\atop
{\rm args}\ \ 
\vec A \to \vec C,  
\vec C \to \vec A,  
\vec q \to -\vec q \, } 
}
\ \ \ \ .  
\hfill
\end{equation}

One may show that the 
post 
and 
prior 
results for the scattering amplitude are
equal 
provided that the external wavefunctions are eigenfunctions of the
Hamiltonian \cite{Schiff}. 
Swanson \cite{S_annphys} 
shows an example of convergence of post and prior results
for meson-meson scattering amplitudes derived from quark Born diagrams as the 
external wavefunctions approach exact Hamiltonian eigenstates.
Of course the Gaussian wavefunctions we use to derive our analytical
results are only
approximations to
the eigenfunctions of the full OGE plus linear Hamiltonian, so in general
we can expect to find a post-prior discrepancy. In this study we actually find such a discrepancy
only in part of the tensor interaction in
PsV scattering, for which we
take the mean
of the two results,
\begin{equation}
T_{fi}(AB\to CD) = 
{1\over 2} 
\bigg(
T_{fi}^{post}(AB\to CD) +
T_{fi}^{prior}(AB\to CD) 
\bigg) \ .
\end{equation}

\newpage
\renewcommand{\theequation}{G\arabic{equation}}
\setcounter{equation}{0}  
\section*{Appendix G: Y$_{LM}$ Expansions and Related Integrals}

It is useful 
in the partial wave decomposition of scattering 
amplitudes 
to expand functions of the sum and difference
momentum transfers $\vec Q_\pm = \vec C \pm \vec A $ 
(here $|\vec A | = |\vec C |$) in spherical
harmonics,
\begin{equation}
f(\vec Q_+^2)
= \sum_{\ell}
f_\ell(\vec A^{\, 2})
\sum_{m} \; 
Y_{\ell m}^*(\Omega_C) 
Y_{\ell m}(\Omega_A)
\ ,
\end{equation}
\begin{equation}
f(\vec Q_-^2)
= \sum_{\ell}
(-1)^\ell f_\ell(\vec A^{\, 2})
\sum_{m} \; 
Y_{\ell m}^*(\Omega_C) 
Y_{\ell m}(\Omega_A)
\ .
\label{fqp_expand}
\end{equation}
This expansion may be inverted to determine the coefficient functions
$\{ f_\ell (\vec A^{\, 2})\} $, 
\begin{displaymath}
f_\ell(\vec A^{\, 2}) 
= 
2\pi 
\int_{-1}^1 
\!\!\!
d\mu \;
P_\ell (\mu)\;  
f(\vec Q_+^2)\;  
\end{displaymath}
\begin{equation}
=
2\pi 
\int_{-1}^1 
\!\!\!
d\mu \;
P_\ell (\mu)\;  
f(2\vec A^{\, 2} (1 + \mu ) )\;  
=
2\pi (-1)^\ell
\int_{-1}^1 
\!\!\!
d\mu \;
P_\ell (\mu)\;  
f(2\vec A^{\, 2} (1 - \mu ) )
\ .
\end{equation}

Many of the scattering amplitudes derived in this paper are proportional to
confluent hypergeometric functions in 
$\vec Q_+^2$ or
$\vec Q_-^2$, 
and their 
partial wave
decomposition requires the integral of a Legendre polynomial times a 
shifted confluent hypergeometric function. 
This integral 
is 
given by
(abbreviating 
${}_1$F$_1(a;c;x)$
as
$f_{a,c}(x)$)
\begin{equation}
{\cal F}^{(\ell)}_{a,c}(x) = 
\int_{-1}^1 
\!\!\!
d\mu \; P_\ell(\mu) \: f_{a,c}(x(1+\mu) ) =
\sum_{m=0}^{\ell} \; c_m^{(\ell)} \; 
{ (a)_{-m-1} \over (c)_{-m-1}} \; 
{ 
\Big(
f_{a-m-1,c-m-1}(2x)  
+ (-1)^{\ell + m + 1}
\Big)
\over
x^{m+1} 
}
\label{teds_excellent_int}
\end{equation}
where
\begin{equation}
c_m^{(\ell)} = 
{ (-1)^m \over 2^m m! } \; 
{ (\ell+m)! \over (\ell-m)! } 
\end{equation}
and the Pochhammer symbol of negative index is
\begin{equation} 
(a)_{-n} = {\Gamma(a - n )\over \Gamma(a)} 
= {1\over  \prod_{k=1}^n (a-k) }\ .
\end{equation}
In numerical evaluations 
it is often useful 
to transform the confluent hypergeometric function in
Eq.(\ref{teds_excellent_int}) to a more rapidly converging
negative-argument form, using
$f_{a,c}(2x)
=  
f_{c-a,c}(-2x)
\; e^{2x}$. 

\newpage
\renewcommand{\theequation}{H\arabic{equation}}
\setcounter{equation}{0}  
\section*{Appendix H: Spin-orbit and Tensor matrix elements}

As noted in App.D, when evaluating phase shifts and inelasticities it is useful
to determine T-matrix elements between 
$|jls\rangle$ states. In the PsV system there are spin-orbit and tensor 
contributions to the T-matrix, and determination of the $j,l,s$-basis matrix
elements of these terms is a complicated problem in angular analysis. Here
we show how these matrix elements may be evaluated.

First consider the spin-orbit terms in the $\pi\rho$ T-matrix, 
Eq.(\ref{pirhoT}). The generic term is of the form

\begin{equation}
T_{fi} =
f(\vec Q_+^2)\;
[
\, i \vec S_\rho \cdot  ( \vec A \times \vec C \, )
] \ ,
\label{Tfiso}
\end{equation}
where $\vec Q_+ = \vec C + \vec A$. (Additional dependence on the rotational
scalar
$\vec A^{\, 2} =
\vec C^{\, 2}$ is a trivial modification of this
angular decomposition.)
To proceed, we expand 
$
f(\vec Q_+^2)
$
in spherical harmonics, as in
Eq.(G1);
\begin{equation}
f(\vec Q_+^2)
= \sum_{\ell}
f_\ell(\vec A^{\, 2})
\sum_{m} \; 
Y_{\ell m}^*(\Omega_C) 
Y_{\ell m}(\Omega_A)
\label{fexpand}
\end{equation}
and introduce spherical components for the spin and momentum vectors,
\begin{equation}                             
\langle 1 s_z' | S_\mu | 1 s_z \rangle =
-\sqrt{2}\;  
\langle 
1 s_z' 
| 
1 \mu , 1 s_z 
\rangle \ ,
\end{equation}
\begin{equation}
i(\vec A \times \vec C )_{\mu} =
{4\sqrt{2}\; \pi \over 3}\, A^2 
\sum_{\mu'\mu''} \; 
\langle 
1 \mu', 1 \mu'' 
| 
1 \mu 
\rangle
\;
Y_{1 \mu' }(\Omega_A) \;
Y_{1 \mu'' }(\Omega_C) 
\end{equation}
and the usual state vector expansion,
\begin{equation}
|jm, ls \; (PsV)\, \rangle = 
\sum_{\mu,s_z} \; 
\langle 
l\mu, 1s_z 
| 
jm 
\rangle\;
Y_{l\mu}(\Omega_A) \;|1s_z \rangle \ .
\end{equation}
With these substitutions one may determine the 
$\pi\rho$
$
\langle j l' s |
T_{fi}
|j l s  
\rangle 
$
matrix elements 
(analogous to the
spinless matrix element $T_l$ of Eq.(D7))
for the spin-orbit term (H1). 
The result involves 
a sum over a product
of six Clebsch-Gordon coefficients, and can be written as the product
of two Wigner 
$ ( 3j ) $ symbols 
and two
$ \{ 6j\} $ symbols, 

\begin{displaymath}
\langle j l' s |
T_{fi}
|j l s  
\rangle 
= 
(-1)^{j+1} 
6{\vec A\, }^2 \, 
\sum_{\ell}
(-1)^{\ell} 
f_\ell \;
(2\ell + 1)\,
\sqrt{ (2l+1)(2l'+1) }\;
\ \  \  \  \  \  \ 
\end{displaymath}
\begin{equation}
\cdot \left(
\begin{array}{ccc}
1 & l & \ell  \\
0 & 0 & 0     
\end{array}
\right)
\left(
\begin{array}{ccc}
1 & l' & \ell  \\
0 & 0 & 0      \\
\end{array}
\right)
\left\{
\begin{array}{ccc}
1 & 1 & 1     \\
l & l' & \ell \\   
\end{array}
\right\}
\left\{
\begin{array}{ccc}
1 & 1 & 1     \\
l & l' & j    \\   
\end{array}
\right\} \ .
\end{equation}
The constraints of the
$ ( 3j ) $ 
and
$ \{ 6j \}  $ 
symbols 
force this matrix element to be diagonal in $l,l'$, and imply that the 
only radial
components of $f(\vec Q_+^{\, 2})$ 
in Eq.(\ref{fexpand}) that contribute are $f_{\ell = l\pm 1 }$.
Substitution of the explicit $ ( 3j ) $ and
$ \{ 6j\} $ symbols
gives our final result for 
$PsV$
matrix elements 
of 
spin-orbit (H1) type,
\begin{equation}
\langle j l' s  |
\;
T_{fi}
\;
|j l s \rangle
=
\delta_{ll'}\;
{
[ j(j+1) - l(l+1) - 2]  
\over 2 (2l+1) }\,
{\vec A\, }^2 
\bigg( f_{l-1} - \, f_{l+1} \bigg) \ .
\end{equation}
This result has the overall 
$\langle \vec {\rm L} \cdot \vec {\rm S}\, \rangle $
dependence that one would expect from a 
spin-orbit force.

We may similarly evaluate the matrix elements of the tensor terms
in Eq.(16). It suffices to consider the two cases

\begin{equation}
T_{fi}^{(t1)} =
f(\vec Q_-^2)\;
\Big(
[
\, \vec S_\rho \cdot \vec A 
\, \vec S_\rho \cdot \vec A 
-{2\over 3}
{\vec A\, }^2
]
+
[
\, \vec S_\rho \cdot \vec C 
\, \vec S_\rho \cdot \vec C 
-{2\over 3}
{\vec C\, }^2
]
\Big)
\end{equation}
and
\begin{equation}
T_{fi}^{(t2)} =
f(\vec Q_-^2)\;
[
\, \vec S_\rho \cdot \vec A 
\, \vec S_\rho \cdot \vec C 
-{2\over 3}
\vec A \cdot \vec C \, 
] \ .
\end{equation}
Both tensor matrix elements have $l\neq l'$ contributions, unlike the other 
interactions we have considered. The general results in terms of Wigner
$( 3j )$ and $\{ 6j \} $ symbols are
\begin{displaymath}
\langle j l' s |
T_{fi}^{(t1)}
|j l s  
\rangle 
= 
(-1)^{j+l+1} 
{2^{1/2} \cdot 5^{1/2} \over 3^{1/2} }\, 
{\vec A\, }^2 \, 
\Big( f_l + f_{l'} \Big) \;
\sqrt{ (2l+1)(2l'+1) }\;
\ \  \  \  \  \  \ 
\end{displaymath}
\begin{equation}
\cdot \left(
\begin{array}{ccc}
2 & l & l'  \\
0 & 0 & 0     
\end{array}
\right)
\left\{
\begin{array}{ccc}
1 & 1 & 2     \\
l & l' & j \\   
\end{array}
\right\}
\end{equation}
and

\begin{displaymath}
\langle j l' s |
T_{fi}^{(t2)}
|j l s  
\rangle 
= 
(-1)^{j+1} 
5{\vec A\, }^2 \, 
\sum_{\ell}
f_\ell \;
(2\ell + 1)\,
\sqrt{ (2l+1)(2l'+1) }\;
\ \  \  \  \  \  \ 
\end{displaymath}
\begin{equation}
\cdot \left(
\begin{array}{ccc}
1 & l & \ell  \\
0 & 0 & 0     
\end{array}
\right)
\left(
\begin{array}{ccc}
1 & l' & \ell  \\
0 & 0 & 0      \\
\end{array}
\right)
\left\{
\begin{array}{ccc}
1 & 1 & 2     \\
l & l' & \ell \\   
\end{array}
\right\}
\left\{
\begin{array}{ccc}
1 & 1 & 2     \\
l & l' & j    \\   
\end{array}
\right\} \ .
\end{equation}
Substitution for the $(3j)$ and $\{ 6j \}$ symbols gives the results quoted
in Eqs.(23-25) in the text.

\newpage

\begin{figure}
\label{fig_1}
\epsfxsize=4truein\epsffile{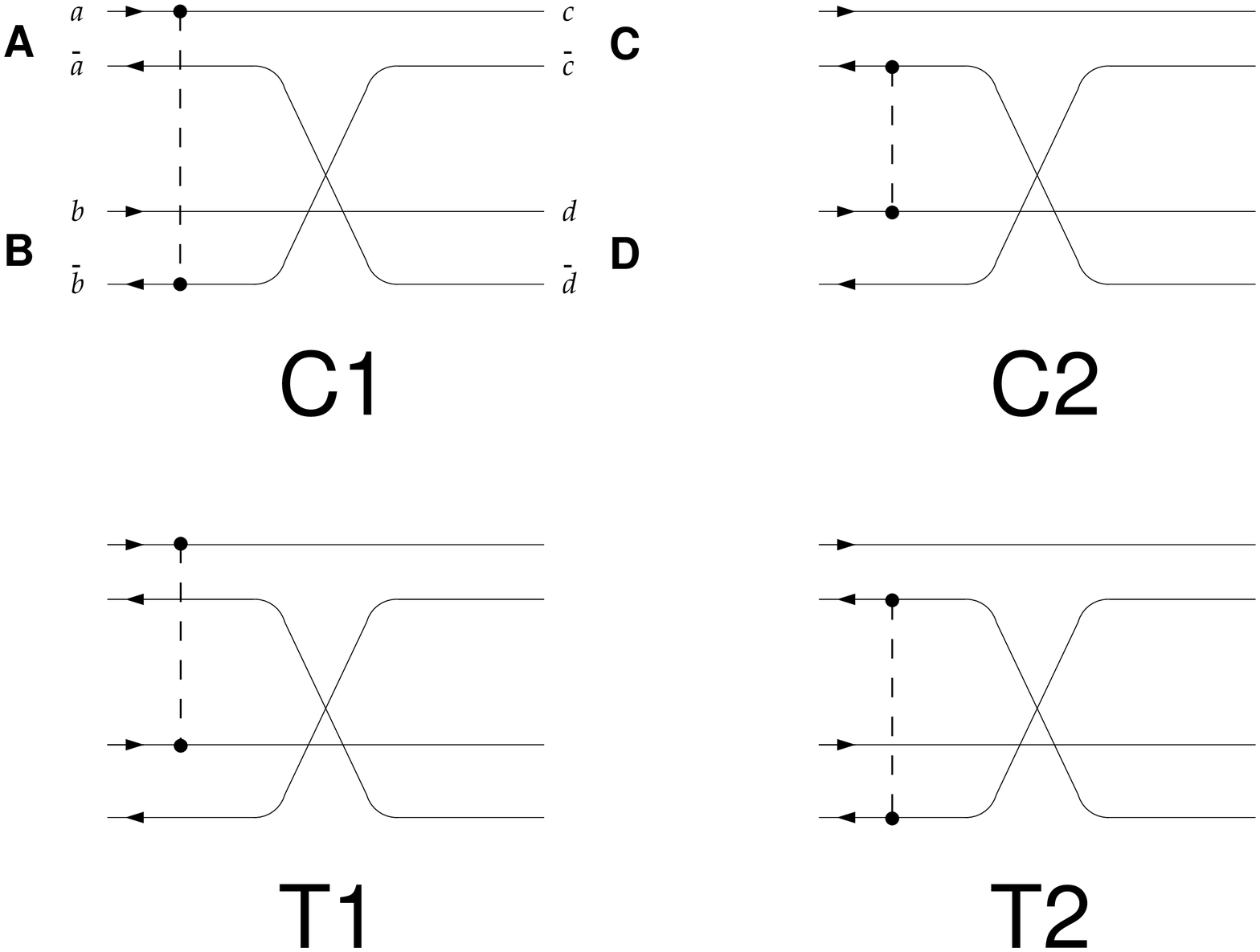}
{Fig.1. The four quark interchange meson-meson scattering diagrams.}
\end{figure}

\begin{figure}
\label{fig_2}
\epsfxsize=4truein
\epsffile{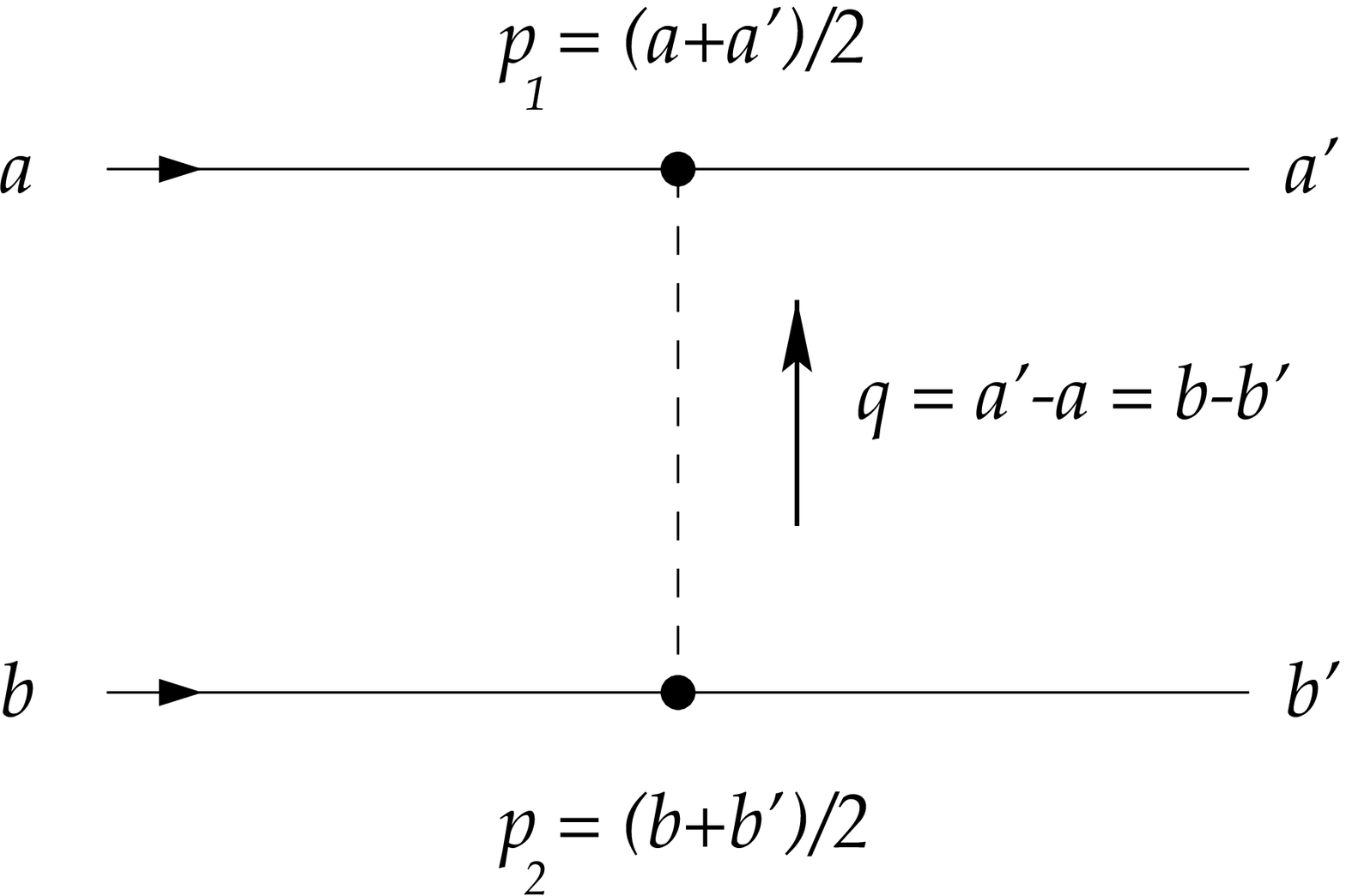}
{Fig.2. The quark-quark T-matrix, showing 
three-momentum definitions.}
\end{figure}

\begin{figure}
\label{fig_3}
\epsfxsize=4.5truein\epsffile{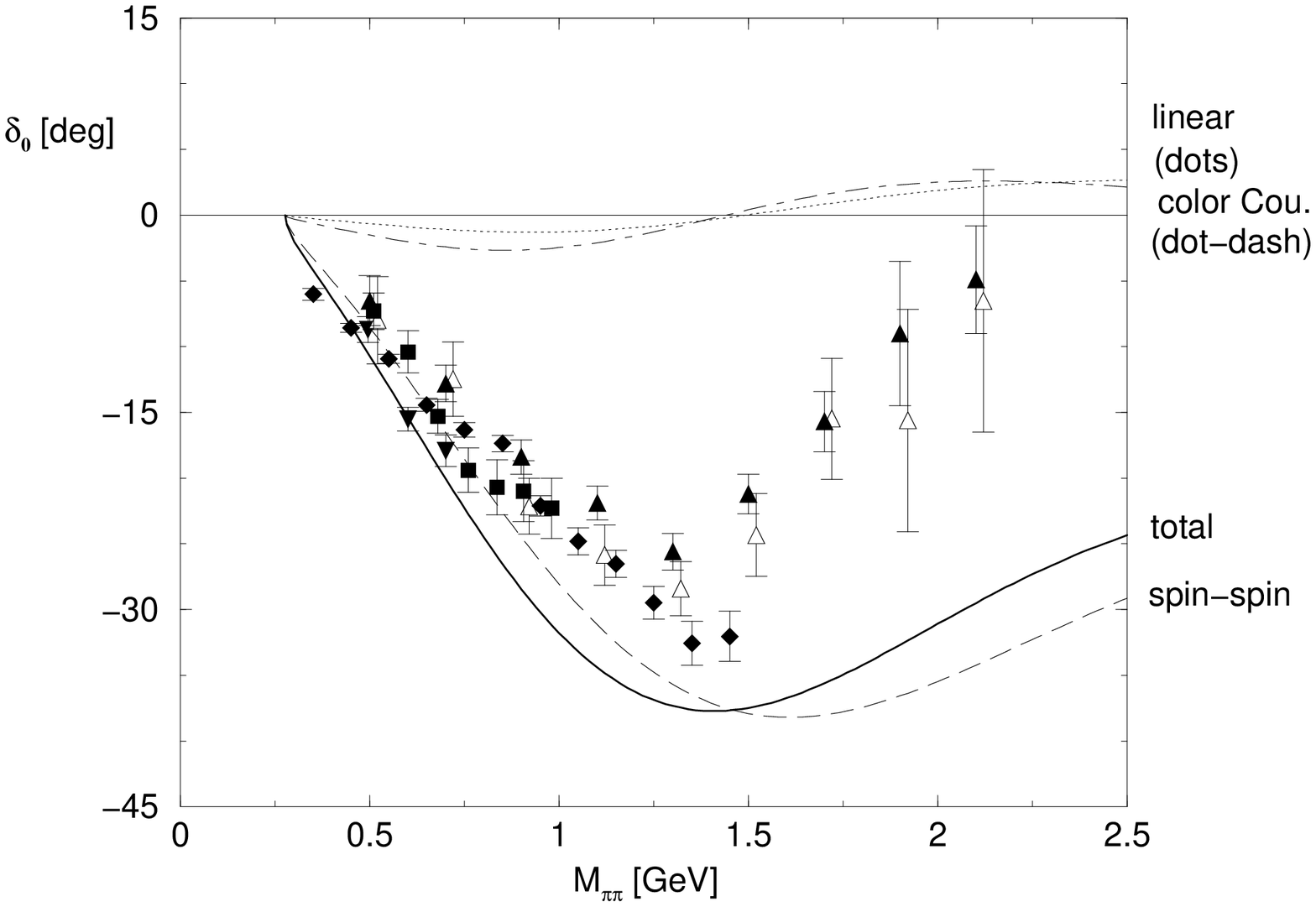}
{Fig.3. Theoretical contributions to
the I=2 $\pi\pi$ S-wave phase shift, 
Eq.(6),
with SHO wavefunctions. The
experimental phase shifts of 
Colton {\it et al.}\cite{Colton} (down triangles),
Durusoy {\it et al.}\cite{Durusoy} (up triangles, two extrapolations),
Hoogland {\it et al.}\cite{Hoogland} (set B, diamonds), 
and 
Losty {\it et al.}\cite{Losty} (squares) 
are shown.}
\end{figure}

\begin{figure}
\label{fig_4}
\epsfxsize=4.5truein\epsffile{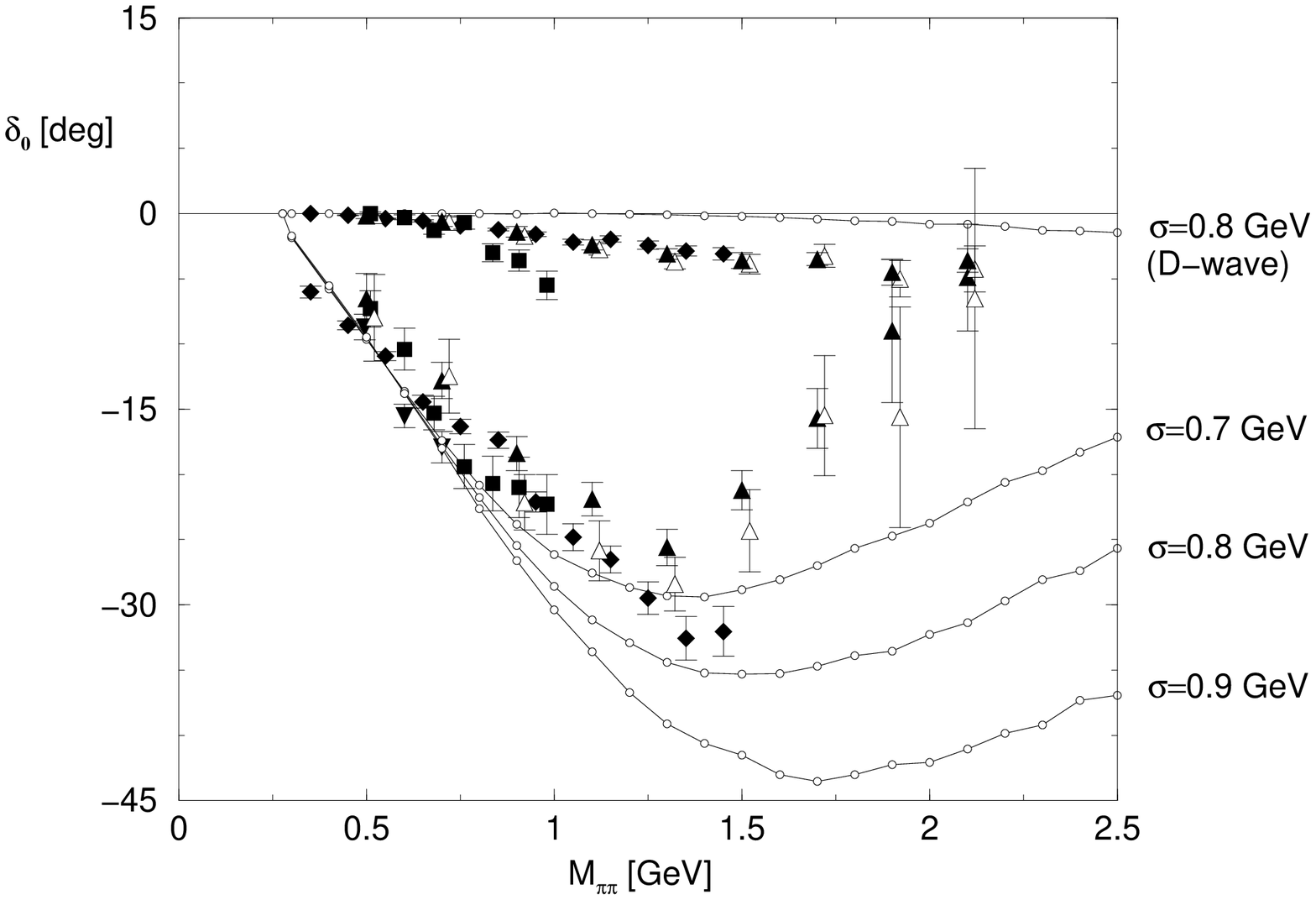}
{Fig.4. Numerically evaluated I=2 $\pi\pi$ S- and D-wave phase shifts 
with Coulomb plus linear plus hyperfine wavefunctions (lines), 
compared to 
experimental phase shifts (symbols as in Fig.3).
}
\end{figure}

\begin{figure}
\label{fig_5}
\epsfxsize=5truein\epsffile{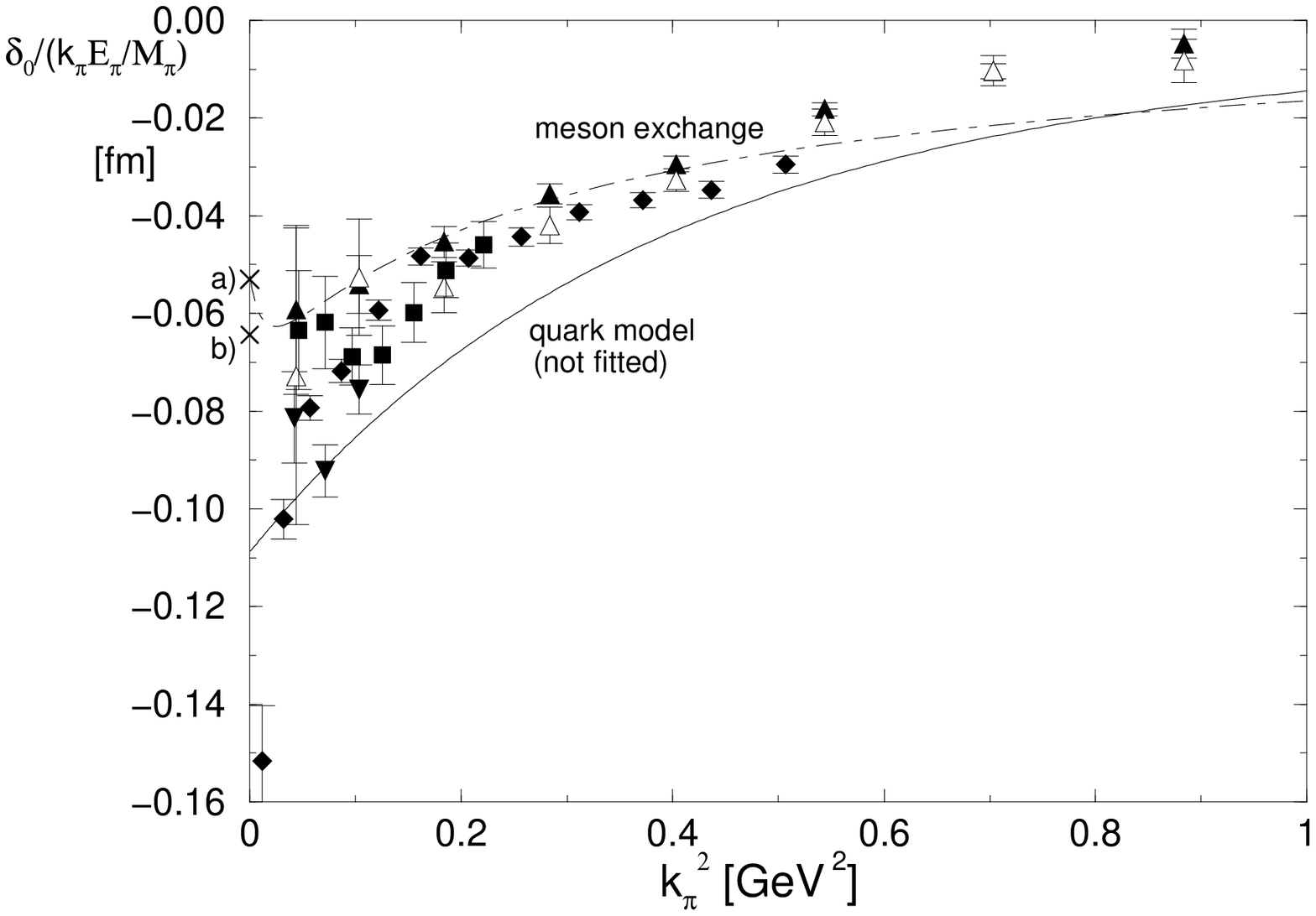}
{Fig.5. A ``generalized specific heat plot" of the
I=2 $\pi\pi$ S-wave phase shift. 
The data 
of Fig.3 is shown together with $a_0^{I=2}$
predictions:  a) LGT \cite{LGT_pipi}, Roy Eqs.\cite{Ana},
b) PCAC \cite{Wein}, $\chi$PT \cite{Don}.
Meson exchange \cite{mesex,FS} and quark model (Eq.(6)) predictions
are also shown. 
}
\end{figure}

\begin{figure}
\label{fig_6}
\epsfxsize=5truein\epsffile{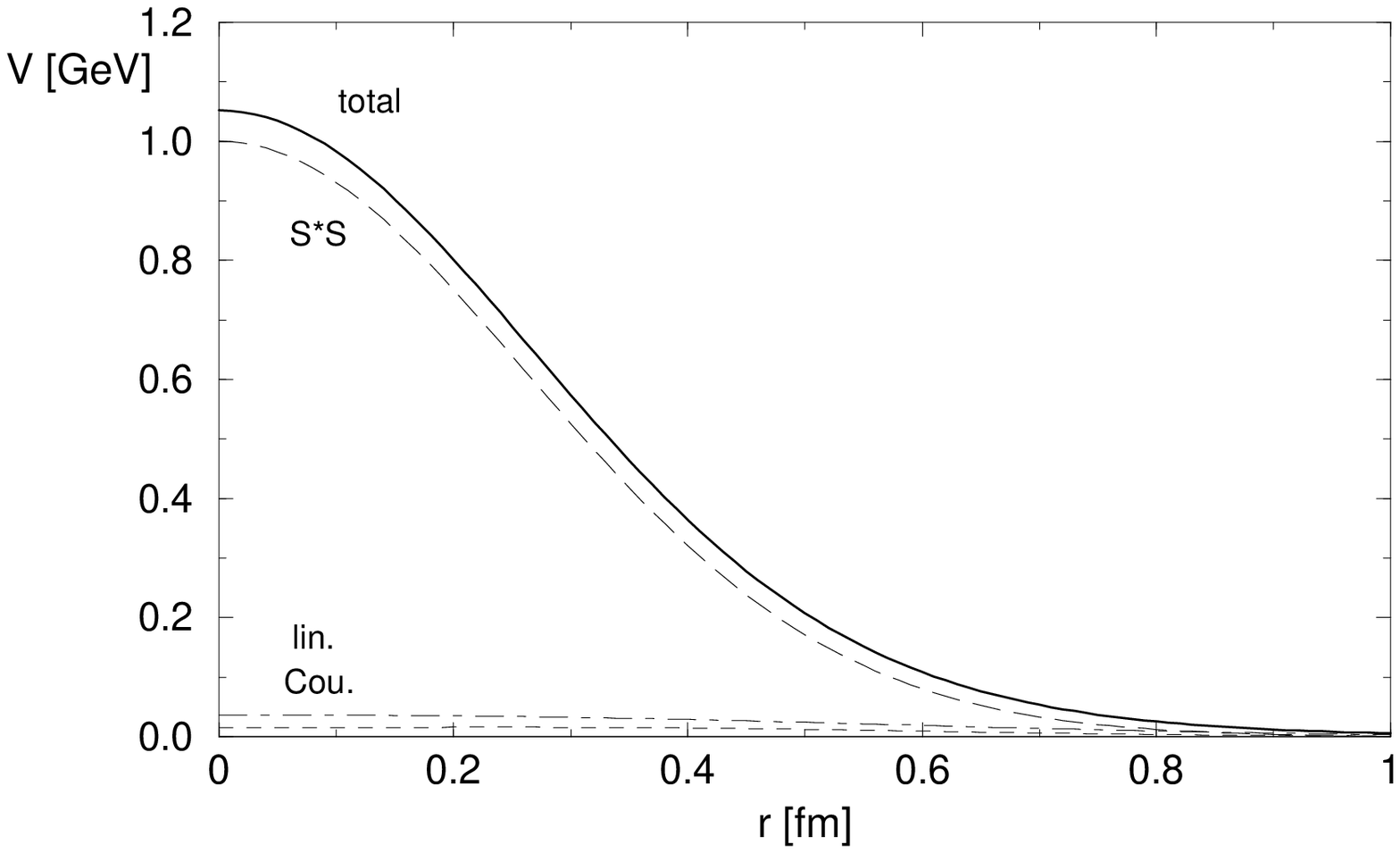}
{Fig.6. The low energy I=2 $\pi\pi$ S-wave potential,
Eq.(13).
}
\end{figure}

\begin{figure}
\label{fig_7}
\epsfxsize=5truein\epsffile{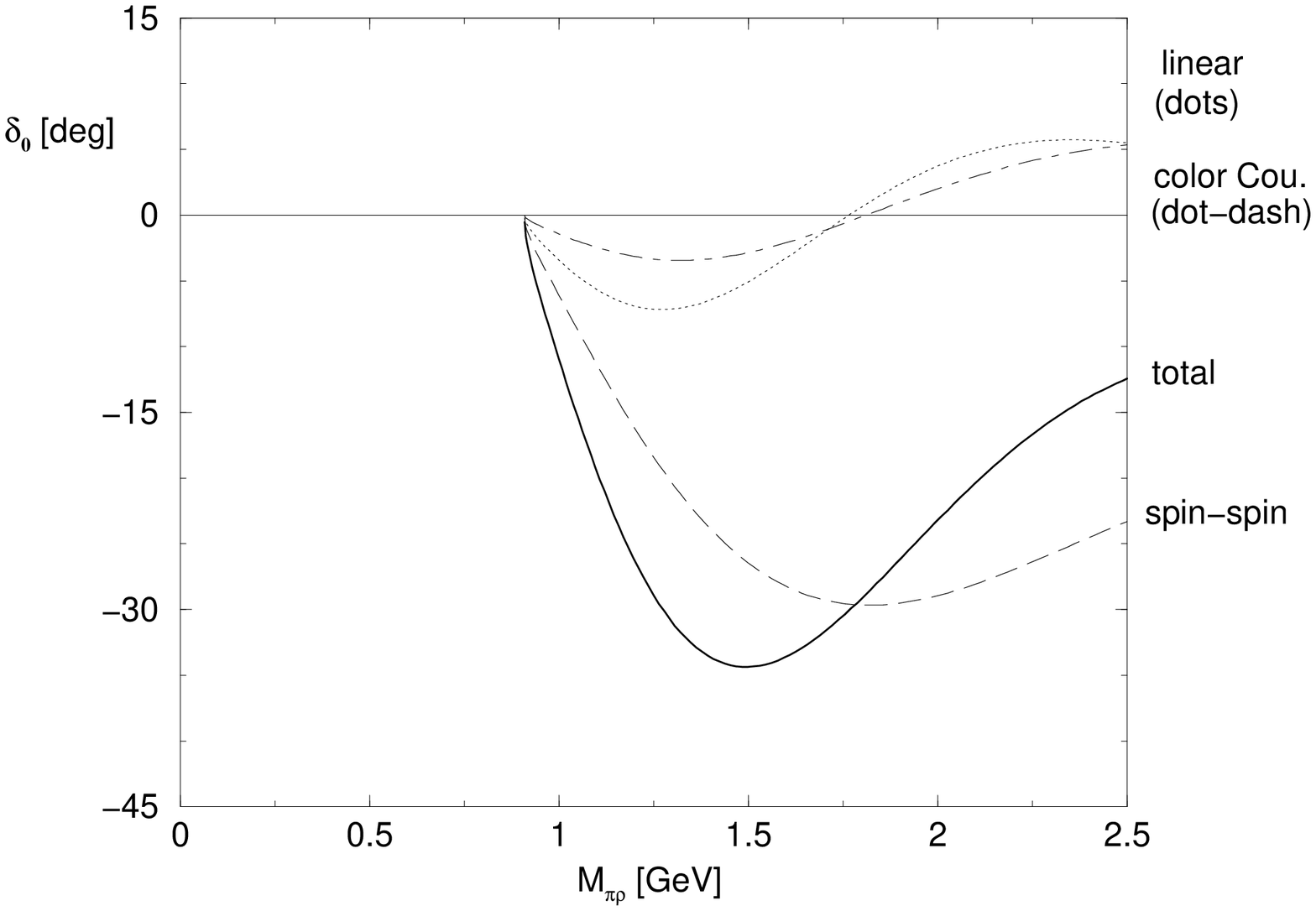}
{Fig.7. The theoretical I=2 $\pi\rho$ S-wave phase shift 
with SHO 
\\
wavefunctions,
Eq.(28).
}
\end{figure}

\begin{figure}
\label{fig_8}
\epsfxsize=5truein\epsffile{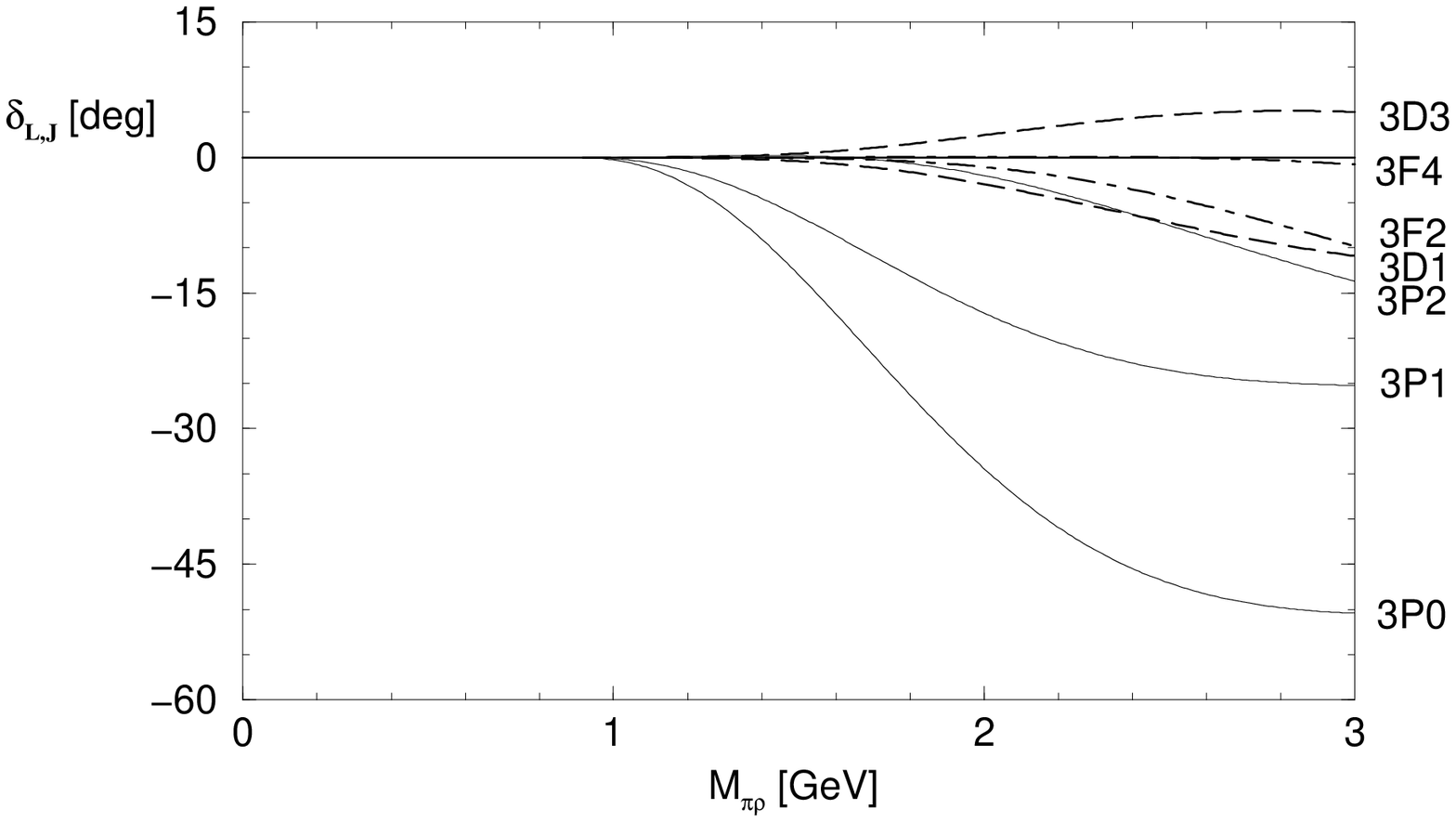}
{Fig.8. Theoretical I=2 $\pi\rho$ phase shifts in P-, D- and F- waves.
\\
$^3$P$_2$, 
$^3$P$_1$, 
$^3$P$_0$ and
J=L$\pm 1$ phase shifts are shown.} 
\end{figure}

\begin{figure}
\label{fig_9}
\epsfxsize=5truein\epsffile{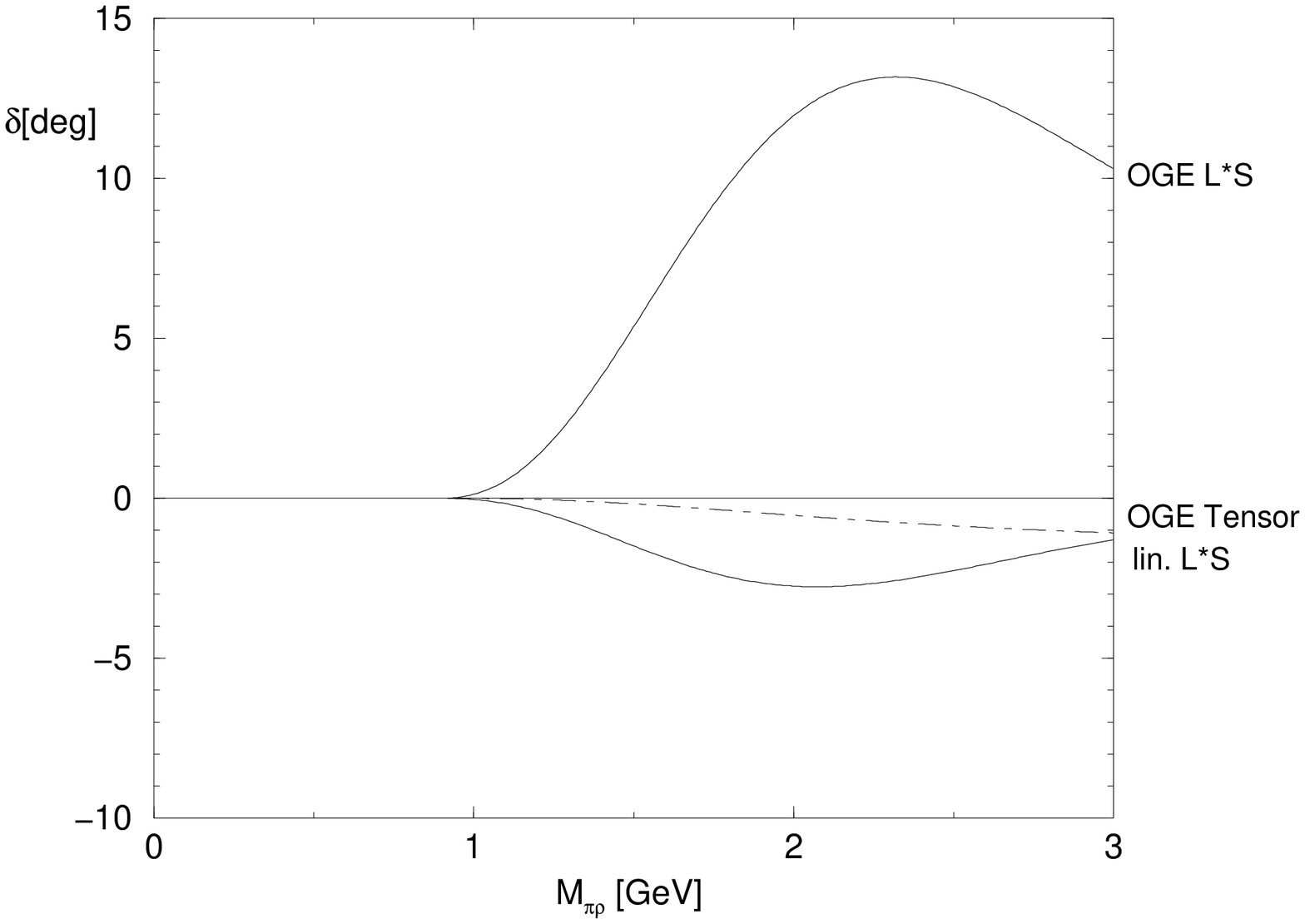}
{Fig.9. 
OGE spin-orbit, confining spin-orbit and OGE tensor\\
contributions to the $^3$P$_2$ I=2 $\pi\rho$ 
phase shift.} 
\end{figure}

\begin{figure}
\label{fig_10}
\epsfxsize=5truein\epsffile{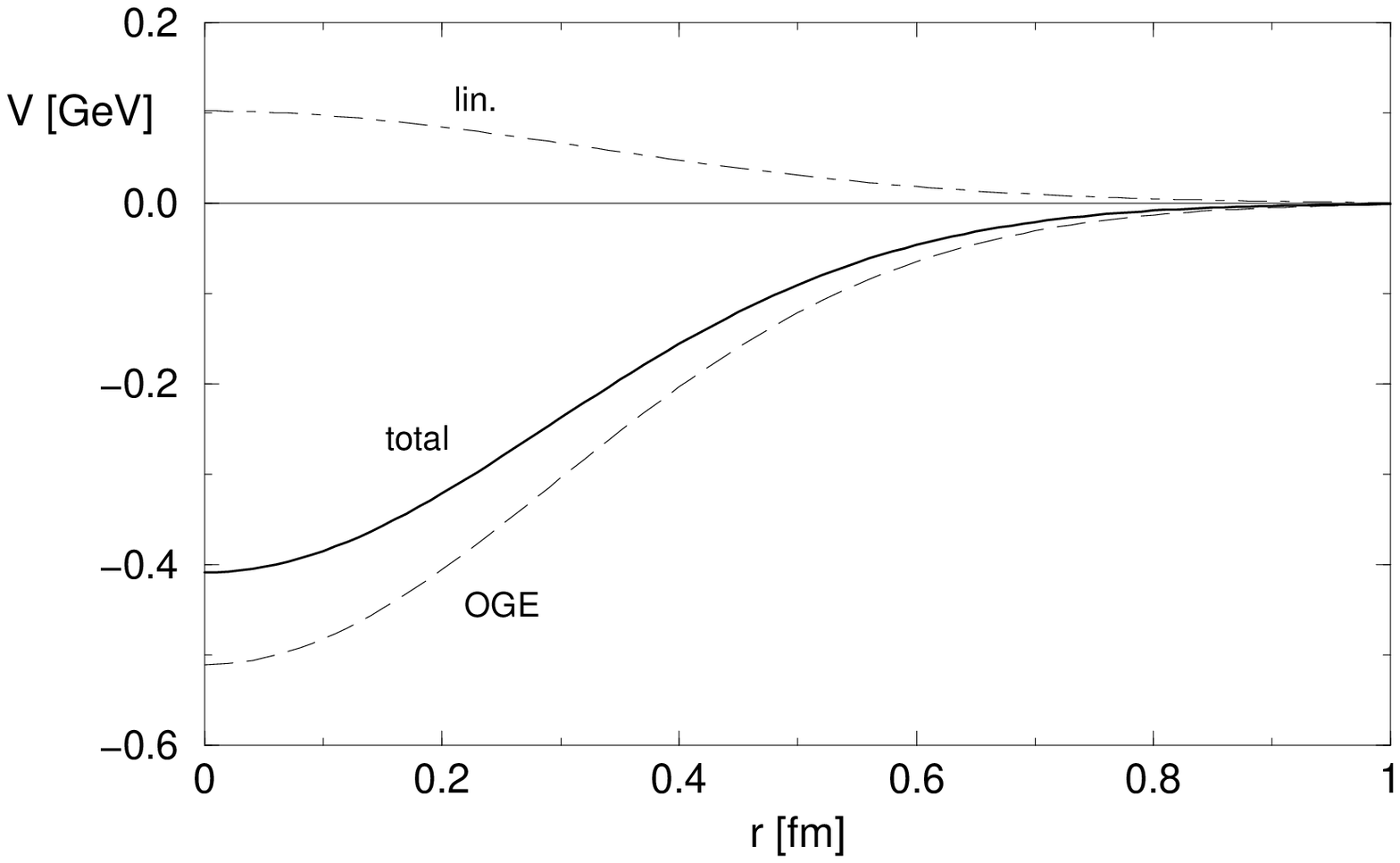}
{Fig.10. 
Spin-orbit potentials in the I=2 $\pi\rho$ $^3$P$_2$ channel,
Eq.(29).}
\end{figure}

\end{document}